\documentclass{article}

\usepackage{arxiv}

\usepackage[utf8]{inputenc} 
\usepackage[T1]{fontenc}    
\usepackage{hyperref}       
\usepackage{url}            
\usepackage{booktabs}       
\usepackage{amsfonts}       
\usepackage{nicefrac}       
\usepackage{microtype}      
\usepackage{lipsum}		
\usepackage{graphicx}
\usepackage{natbib}
\usepackage{doi}

\usepackage{algorithm}
\usepackage{algpseudocode}
\usepackage{float}
\usepackage{amsmath} 
\usepackage{multirow}
\usepackage{subfig}

\algnewcommand\algorithmicforeach{\textbf{for each}}
\algdef{S}[FOR]{ForEach}[1]{\algorithmicforeach\ #1\ \algorithmicdo}

\title{Towards spatiotemporal integration of bus transit with data-driven approaches}

\author{ \href{https://orcid.org/0000-0001-8779-6213}{\includegraphics[scale=0.06]{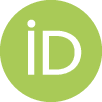}\hspace{1mm}Júlio Borges}\\
	Universidade Tecnológica Federal do Paraná \\
	Curitiba, Brasil \\
	\texttt{julio.2018@alunos.utfpr.edu.br} \\
	\And
	\href{https://orcid.org/0000-0001-9798-578X}{\includegraphics[scale=0.06]{orcid.png}\hspace{1mm}Altieris M. Peixoto} \\
	Universidade Tecnológica Federal do Paraná\\
	Curitiba, Brasil \\
	\texttt{altieris.marcelino@gmail.com} \\
        \And
	\href{https://orcid.org/0000-0001-6994-8076}{\includegraphics[scale=0.06]{orcid.png}\hspace{1mm}Thiago H. Silva} \\
	Universidade Tecnológica Federal do Paraná\\
	Curitiba, Brasil \\
	\texttt{thiagoh@utfpr.edu.br} 
        \And
	\href{https://orcid.org/0000-0002-0182-7128}{\includegraphics[scale=0.06]{orcid.png}\hspace{1mm}Anelise Munaretto} \\
	Universidade Tecnológica Federal do Paraná\\
	Curitiba, Brasil \\
	\texttt{anelise@utfpr.edu.br} 
        \And
	\href{https://orcid.org/0000-0001-6483-4694}{\includegraphics[scale=0.06]{orcid.png}\hspace{1mm}Ricardo L\"{u}ders} \\
	Universidade Tecnológica Federal do Paraná\\
	Curitiba, Brasil \\
	\texttt{luders@utfpr.edu.br} 
}

\renewcommand{\shorttitle}{Towards spatiotemporal integration of bus transit with data-driven approaches}

\hypersetup{
pdftitle={Towards spatiotemporal integration of bus transit with data-driven approaches},
pdfsubject={cs.CE},
pdfauthor={Júlio Borges, Altieris M. Peixoto, Thiago Silva, Anelise Munaretto, Ricardo L\"{u}ders},
pdfkeywords={bus transit network,  spatiotemporal integration, data-driven model, urban computing, smart city},
}

\begin{document}


\maketitle

\begin{abstract}
This study aims to propose an approach for spatiotemporal integration of bus transit, which enables users to change bus lines by paying a single fare. This could increase bus transit efficiency and, consequently, help to make this mode of transportation more attractive. Usually, this strategy is allowed for a few hours in a non-restricted area; thus, certain walking distance areas behave like ``virtual terminals.'' For that, two data-driven algorithms are proposed in this work. First, a new algorithm for detecting itineraries based on bus GPS data and the bus stop location. The proposed algorithm's results show that  $90$\% of the database detected valid itineraries by excluding invalid markings and adding times at missing bus stops through temporal interpolation. Second, this study proposes a bus stop clustering algorithm to define suitable areas for these virtual terminals where it would be possible to make bus transfers outside the physical terminals.
Using real-world origin-destination trips, the bus network, including clusters, can reduce traveled distances by up to $50$\%, 
 making twice as many connections on average.
\end{abstract}



\section{Introduction}

The collective public transport service has been little expanded and modernized over the last few decades, compared to private individual transport, which currently consumes the largest available urban road space. \cite{Motta2013} show historical conflicts and contradictions at the origin of the problems affecting collective public transport systems throughout Brazil.

Aiming to revert this, attract new users, and improve the quality of the service, several companies that manage collective public transport and researchers in this area are using vehicle GPS data more frequently by providing an efficient and accurate way to track the position of vehicles at a given time  \citep{Bona2016, Curzel2019, Santin2020b,gubert2023strategies}. In this way, public transport can be investigated from different perspectives, examining the system's dynamic behavior, measuring the efficiency of services, identifying movement patterns, integration capacity, peak hours, etc.

In analyzing the operation of a public transport network, it is generally necessary to identify the instants of time in which the bus passes at the stops, whether or not the bus stops. In general, this identification is done by a map-matching algorithm crossing the geolocation information of the bus with the location of the bus stop. A suitable method for this task was employed in the work of \citep{Martins2022}. The method presents a solution for detecting bus stops, even on two-way roads, correcting GPS inaccuracies, and identifying the exact time of passage at a bus stop.

However, working with real GPS data in practice can become very difficult because vehicles fail in GPS communication, generating gaps that can affect the analysis results if not treated properly. Another problem is the need to cross-reference scheduled timetables with GPS information, as outdated timetables generate inconsistencies, such as the cases reported by \cite{Martins2022}. Thus, researchers are often forced to look for periods in which there is consistency in logs and tables. Nevertheless, this limitation confines the analysis to specific time intervals and conditions during which the data is optimal.

The itinerary detection problem, as defined in this article, goes further, as it consists of matching the GPS logs of the movement of a bus with its respective itinerary and schedule, which is defined in the bus operation planning. This problem is affected by communication failures in sending GPS data, making it impossible to detect the bus itinerary correctly. An approach for detecting itineraries was employed in the work of \citep{Peixoto2020} using the schedule table for a bus line. In this case, both the starting point departure time and the scheduled arrival time at the endpoint must be known. In addition, logs of bus movements are needed. Although this algorithm identifies most itineraries, several logs from bus monitoring are discarded, as they do not appear associated with any itinerary. Unlike the previous work, the algorithm proposed in this article does not use the scheduled bus line schedule.

In this context, this article aims to develop a spatiotemporal integration of bus transit using data-driven strategies. In our previous work \citep{courb23Julio}, an algorithm was introduced capable of detecting the itinerary of a bus in operation using data from its geolocation without the need to use information from the scheduled timetables of each bus line. Furthermore, the proposed algorithm adds missing data due to communication failures by interpolating known time values.

The main contributions of this paper are as follows:
\begin{itemize}
    \item Bus itinerary detection algorithm for building itineraries with spatiotemporal information based on GPS bus monitoring;
    \item Analysis of the impact of the interpolation of GPS data on the different types of lines of the transport system;
    \item Full database assessment using real GPS bus data and temporal evaluation of bus service;
    \item Bus stop clustering algorithm for grouping nearby stops where connections can be made between bus lines with a single fare;
    \item Evaluation of the bus stop clustering algorithm and the bus service synchronization;
    \item Evaluation of the spatiotemporal integration impact on bus trips in terms of distance traveled and number of transfers between bus lines using real-world origin-destination trips.
\end{itemize}

The article's structure is as follows: In \textbf{Section~\ref{sec:trab_relacionados}}, an overview of related works is presented. \textbf{Section~\ref{sec:algoritmo}} provides a detailed explanation of the proposed algorithms. The outcomes of these algorithms are discussed in \textbf{Section~\ref{sec:resultados}}, followed by the conclusion in \textbf{Section~\ref{sec:conclusao}}.

\section{Related Works} \label{sec:trab_relacionados}

\subsection{Public Transport Mobility Understanding}
The popularization of very low-cost embedded systems technologies such as Arduino\footnote{https://www.arduino.cc/}, Raspberry Pi\footnote{https://www.raspberrypi.com/} and ESP32\footnote{https: //www.espressif.com/en/products/socs/esp32} enabled the growth of research involving Internet of Things (IoT) and public transport. In particular, the installation of this equipment allows the creation of a wide range of applications with the potential to promote service improvement and efficiency. The following studies of \cite{Sridevi2017}, \cite{Hakeem2022}, and \cite{Desai2022} exemplify recent applications in which onboard equipment collects the GPS trajectory of buses and centralizes it on a server. This data can be used in various transportation system management applications. A literature review on the bus trajectory data application can be seen in \citep{War2022}. Aspects such as data sources and methods of Big Data and IoT in mass public transport are described in \citep{Welch2019}. Further discussion of using bus GPS trajectory data is provided in \citep{Singla2015}.

Concerning the Integrated Transport Network (ITN), several studies such as \citep{Bona2016, Curzel2019, Santin2020b,Rosa2020, gubert2023strategies}, used open public transport data, such as vehicle GPS trajectory data, timetables, and itineraries, in the creation of models that allowed the expansion of the understanding of the characteristics and behaviors of the transport network, to promote the improvement of the efficiency of the service. Other studies on urban mobility, transport networks, and computational models that use similar data are performed by \cite{Rodrigues2017}, \cite{Wehmuth2018}, \cite{Maduako2019c}, \cite{Sadeghian2021}, and \cite{Li2022}. These models provide many efficiency measures for transport services, helping to identify opportunities for important improvements such as cost reduction.

\subsection{Public Transport Data Quality}

Another important issue is the data quality. Cleaning the raw data must reduce inconsistencies for the models to work properly. For example, \cite{Martins2022} pointed out several problems present in real GPS data, and therefore, they developed a model of map matching. This model can be used for: i) vehicle stop detection between nearby bus stops on a two-way street, in which one point serves both directions (``going'' and  ``returning'') from the bus line; ii) GPS inaccuracies; and iii) the vehicle's exact time of passage at a bus stop. The problem of detecting bus stops was also addressed in the work of \cite{Peixoto2020}, where the vehicle's itinerary was also employed using data from timetables. Other authors also faced the challenge of detecting compatibility between the trajectory of buses and their respective itineraries, such as \citep{Yin2014, Queiroz2019, Chawuthai2023}. In these works, the main objective of detection is to identify whether or not a bus GPS trajectory is in accordance with the planned itinerary to signal any inconsistency. In the work of \cite{Gallotti2015}, inconsistencies in vehicle stop times were corrected using a temporal interpolation method, but no measure of interpolation error was presented.

The algorithm proposed in the present article fills some gaps, identifying and correcting inconsistencies in GPS trajectories according to the itinerary. Furthermore, a method to measure the interpolation error is presented. Therefore, the proposed approach treats the itinerary detection problem and the temporal interpolation by reducing the inconsistencies in the raw data and providing a reliable database for new applications.

\subsection{Strategies to Increase Public Transport Efficiency}

Several studies concentrate efforts on understanding public transportation-specific characteristics, which could be useful for many tasks, such as improving robustness, tackling problems, promoting better use by users, or inspiring better optimization strategies \citep{Zhang2021,Li2022,War2022,Park2020,Maduako2019c,Gallotti2015b}.

Another group of studies closer to our work proposes strategies to improve public transport efficiency \citep{yu2020policy,mulerikkal2022performance,ma2019bus,zhao2023developing,ma2022multi}. For instance, the authors in \cite{yu2020policy} present policy zoning with different management strategies to make the dockless bicycle-sharing service a more practical travel mode for linking the metro system. By studying real usage data, the authors show that metro stations in Shanghai are classified into four clusters with different characteristics, demand patterns, and operation performance; thus, corresponding policy recommendations are proposed. 

The work presented in 
\cite{mulerikkal2022performance} proposes a deep neural network model for predicting subway passenger flow. It relies on spatiotemporal data provided by automated  
card systems from subway usage. Passenger flow prediction is central to efficient transportation management. Another work~\cite{ma2019bus} presents a method for forecasting bus journey duration by integrating real-time data from taxis and buses, which can intelligently segment bus routes into dwelling and transit portions. Two distinct models were developed to forecast these segments individually, considering various traffic variables. Anticipating public transportation schedules can minimize passenger waiting times and encourage greater use of public transit.

\cite{Caminha2018} presented a data-driven approach that assesses the imbalance between supply and demand in public transit of Fortaleza, a Brazilian city with a population of over $2.5$ million inhabitants. Their methodology considers real GPS and ticketing data of a bus system to build a complex network. This network is used to find places where supply is insufficient or stretches of the network where buses travel almost empty. Their methodological contributions could be used to identify opportunities to improve resource distribution in areas where bus demand is high, but bus supply is low by relocating buses that are being underutilized.

The data-driven approach outlined in this article enhances our knowledge base and methodological toolkit by introducing algorithms and a systematic process for assessing the spatiotemporal integration of public transport networks. This method allows for a comprehensive examination, enabling the identification of opportunities to enhance the utilization of resources, including buses and existing infrastructure.

\section{Spatiotemporal Integration of Bus Transit} \label{sec:algoritmo}

A spatiotemporal integration of bus transit is a strategy that allows users to change bus lines by paying a single fare. Usually, this strategy is allowed for a few hours in a non-restricted area (terminals are restricted areas, for instance). Certain walking distance areas function similarly to ``virtual terminals''. Two goals must be considered: i) detection of bus itineraries and ii) bus stop clustering to define suitable areas for these ``virtual terminals.'' Both objectives are achieved by the two data-driven algorithms proposed in this section.

The central issue is to develop an itinerary detection algorithm independent of the bus schedule table. Initially, it is necessary to differentiate the concepts of ``static network'' and ``dynamic network.'' These concepts were also used in the work of \citep{Peixoto2020}. A static network represents the topology of the transport network, that is, the sequencing of bus stops on a specific line covering all itineraries offered by the service, as considered in \citep{Bona2016}.

Since the static network describes the topology of the bus line and its respective itineraries without including timetables, the dynamic network is formed as a given vehicle reaches the points provided for in its service itinerary. The proposed itinerary detection algorithm is composed of 3 steps:

\begin{itemize}
\item \textbf{step 1}: mark the time a bus passes at bus stops (map matching algorithm).
\item \textbf{step 2}: sequence bus stops according to these time marks (temporal sequencing).
\item \textbf{step 3}: associate a temporal sequence of bus stops to a known itinerary, interpolating and removing marks if necessary (the proposed algorithm).
\end{itemize}

A map-matching algorithm is used at \textbf{step 1}. The algorithm used in this work is based on \citep{Martins2022} algorithm. It calculates the Haversine distance from each vehicle position ($log$) as used in \citep{panigrahi2014computing, lawhead2015learning} to all bus line stops and assigns the $log$ to the closest stop. This way, it is possible to mark the time of passage of a vehicle to each bus line stop. \textbf{Step 2} sorts time marks in ascending order to obtain a temporal sequencing of points. \textbf{Step 3} is accomplished by the algorithm proposed below.

\begin{figure}
\begin{center}
\includegraphics[width=.70\columnwidth]{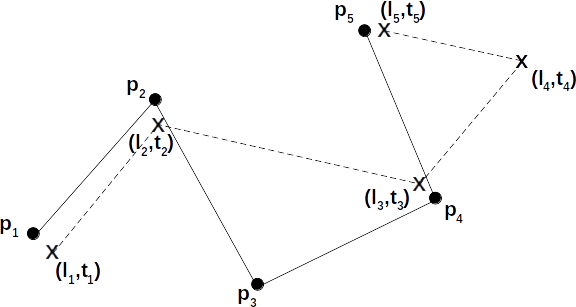}
\caption{Example of an itinerary $iti=(p_1, p_2, p_3, p_4, p_5)$ in full line and a $log=((l_1,t_1), (l_2,t_2), (l_3,t_3), ( l_4,t_4), (l_5,t_5))$ in dashed line.}
\label{fig:iti}
\end{center}
\end{figure}

The itinerary detection algorithm aims to associate a $log$ of events captured by the movement of a specific bus to the sequence of points of their respective bus line registered in the table \texttt{LinePoints}. Thus, it is possible to associate the instant time of passage of the bus at all points on the line. This is illustrated in \textbf{Figure~\ref{fig:iti}}, where a $log$ of events $log=((l_1,t_1), (l_2,t_2), (l_3,t_3), (l_4,t_4), (l_5,t_5))$ will be associated with a itinerary $iti=(p_1, p_2, p_3, p_4, p_5)$, where $t_i$ is the moment when the bus passes at coordinate $l_i$ and $p_i$ is a point on the bus itinerary.

For example, according to \textbf{Figure~\ref{fig:iti}}, there is no record of a bus passing at point $p_3$, and there is a record of a bus passing at a position $l_4$ that does not correspond to any registered point on the line.

The map matching algorithm associates the locations $l_i$ to the respective bus stops by evaluating a measure of spatial proximity between $l_i$ and a point on the itinerary. In the case of \textbf{Figure~\ref{fig:iti}}, the result of map matching is the mapping $map=((p_1,l_1), (p_2,l_2), (p_4,l_3), (-, l_4), (p_5,l_5))$, and no point is associated with location $l_4$ and there is no record of the passage through point $p_3$.

Then, the proposed algorithm detects the itinerary with temporal information $det=((p_1,t_1), (p_2,t_2), (p_3,\hat{t}_3), (p_4,t_3), (p_5,t_5) )$, associating the moment of passage to each point of the bus line. In this case, the time instant $\hat{t}_3=t_2+(t_3-t_2)/2$ is estimated by averaging the times $t_2$ and $t_3$ of points adjacent to $p_3$.

The above result for itinerary detection $det=\{(p_i,t_j)\}$ with temporal information can be generalized to ~\eqref{eq:iti} for $iti$ of dimension $n$ and $log$ of dimension $m$.
\begin{equation}
\label{eq:iti}
\footnotesize
(p_i,t_i) =
\left\{
\begin{array}{ll}
(p_i,t_j) &: (p_i,l_j) \in map; i=1,\dots,n; j=1,\dots,m\\
(p_i,\hat{t}_i) &\textrm{otherwise}\\
\end{array}
\right.
\end{equation}
where $\hat{t}_i=\hat{t}_{i-1}+(t_{k+w}-t_k)/w$ for $k<i<(k+w)$ and $\ hat{t}_k=t_k$, considering the $w-1$ bus stops that were not mapped by map matching between points $(p_k,t_k)$ and $(p_{k+w}, t_{k+w}) \in map$. The above procedure is summarized in \textbf{Algorithm~\ref{alg:det_iti}}.

\begin{algorithm}[h]
\caption{Itinerary detection} \label{alg:det_iti}
\begin{algorithmic}[1]
\begin{footnotesize}
\Require $iti = \{p_i\}$, $1 \le i \le n$; $log = \{(l_j,t_j)\}$, $1 \le j \le m$ // ordered by points and time, respectively
\Ensure $det = \{(p_i,t_i)\}$, $1 \le i \le n$
\State $map \gets \{\}$
\State $time \gets -1$
\ForEach {$p_i \in iti $}
\State{$found \gets False$}
\ForEach {$(l_j,t_j) \in log $}
\If{$(p_i = l_j)$ and $(t_j > time)$}
\State $map \gets map \cup \{(p_i,l_j,t_j)\}$
\State $found \gets True$
\State $time \gets t_j$
\EndIf
\EndFor
\If{$(found = False)$}
\State $map \gets map \cup \{(p_i,None,None)\}$
\EndIf
\EndFor
\State $det \gets \{\}$
\ForEach {$(p_i,l_i,t_i) \in map$}
\If{$(l_i \ne None)$}
\State $det \gets det \cup \{(p_i,t_i)\}$
\Else
\State computes $w$, $\Delta t=(t_{k+w}-t_{k})$ and $\hat{t}_k=t_k$ // missing points between $p_k$ and $p_{k +w}$ acquaintances
\State $\hat{t}_i=\hat{t}_{i-1}+\Delta t/{w}$
\State $det \gets det \cup \{(p_i,\hat{t}_i)\}$
\EndIf
\EndFor
\end{footnotesize}
\end{algorithmic}
\end{algorithm}

The proposed algorithm has as requirements the static network and the dynamic network. That is, it is necessary to provide as input both the structure (or topology) of the transport network, according to the file \texttt{PontosLinha}, as well as the \textit{logs } of GPS of buses from file \texttt{Vehicles}. The advantage over the method proposed by~\cite{Peixoto2020} is that there is no need for a table of scheduled bus lines (files \texttt{TabelaLinhas} and \texttt{TabelaVeículos}).

Given a set of markings with known times $(p_i,t_i)$ and estimated times $(p_i,\hat{t}_i)$, temporal interpolation introduces an estimation error given by $err_i=|t_i-\hat{t }_i|$. This work takes known values from the original data set to obtain the estimation error evaluation results.

The second algorithm deals with bus stop clustering that behaves like virtual terminals  (not constrained to a particular area), where users can change between bus lines with a single fare. They are defined when nearby bus stops are clustered within a $600$ meter radius, which is considered a walking distance suitable for changing bus lines \cite{Peixoto2020}.

Centroids and clusters are computed according to \textbf{Algorithm~\ref{alg:cluster}}. It starts with a descending-order list of candidate centroids ordered by the average number of buses. For each centroid in the head of the list (line 3), all neighbor bus stops within a given distance from the centroid are included in the cluster (line 7). The new cluster is then added to the set of clusters (line 10), and all bus stops of the cluster are removed from the candidates list. It means that clustered bus stops are no longer candidates for another cluster. The algorithm ends when the list of candidates is empty (line 12).

\begin{algorithm}[h]
\caption{Bus stop clustering} \label{alg:cluster}
\begin{algorithmic}[1]
\begin{footnotesize}
\Require$candidates = \{c_i\}$, $1 \le i \le n$; // ordered list of $n$ centroid candidates 
\Statex $bus\_stops = \{b_j\}$, $1 \le j \le m$  // list of bus stops
\Ensure $clusters = \{\{cluster\}_i\}$, $1 \le i \le l$ // set of $l$ clusters 
\State $clusters \gets \{\}$
\While {$candidates \ne \{\}$}
\State $centroid \gets head(candidates)$ // first element of the ordered list
\State $cluster \gets \{centroid\}$
\ForEach {$b_j \in bus\_stops$}
\If{$haversine(centroid, b_j) \le 600$}
\State $cluster \gets cluster \cup \{b_j\}$
\EndIf
\EndFor
\State $clusters \gets clusters \cup \{\{cluster\}\}$
\State $candidates \gets candidates - (candidates \cap cluster)$ // remove clustered bus stops from candidates list 
\EndWhile
\end{footnotesize}
\end{algorithmic}
\end{algorithm}

\section{Results and Discussions} \label{sec:resultados}

The evaluation of our proposal is accomplished using the real bus GPS data.
Data description, itinerary detection, and evaluation of interpolation errors are presented in \textbf{Sections~\ref{sec:dados} to \ref{sec:limitacoes}}. Results for spatiotemporal integration of bus transit are presented in \textbf{Sections~\ref{sec:temporal} to \ref{sec:impact}}.

\subsection{Public Transport Data} \label{sec:dados}

The C3SL repository\footnote{\url{http://dadosabertos.c3sl.ufpr.br/curitibaurbs/}} is recognized as the main source of data for academic applications on public transport in Curitiba. It has been used in several studies.

Geolocated data from bus monitoring are needed to deal with the map matching problem and static information from the transport network and the bus schedule. The Open Data Portal of Curitiba City Hall provides a daily updated database containing data on public transport in Curitiba available via WebService with relevant information such as GTSF Files, Lines, Points, Itineraries, Position of Vehicles, and Tables of Schedules. Data is transferred through files in JSON format through an API. The API data dictionary can be found in technical documentation~\citep{PDA_URBS_API}.

According to the operational data published in \citep{URBS_RIT}, ITN has a fleet of 1,226 vehicles (disregarding the reserve buses) that serve 250 lines, 22 terminals, and 329 tube stations and make, on average, 1,365,615 trips per day useful. These vehicles periodically send their location according to~\citep{PDA_URBS_API}, which is stored in a daily log to be consulted via the API. Because the native service does not offer requests by date, C3SL provides JSON files containing a daily and complete history of ITN operations updated on day 1. An extensive exploratory analysis of these data is given in \citep{vila2016urban}.

The following data files of C3SL are used in the experiments:
\begin{itemize}
\item \texttt{Linhas}: contains code, name, service category, color, and other attributes of all ITN bus lines.
\item \texttt{PontosLinhas}: stores name, code, type, latitude, and longitude of all ITN bus stops, and describes the correct sequence of stops according to bus line itineraries.
\item \texttt{Vehicles}: contains the coordinate history of vehicles in operation. The GPS position of a vehicle with date and time is sampled every 20 seconds on average.
\item \texttt{TabelaLinhas}: stores bus line timetables at stops;
most stops do not have timetable information.
\item \texttt{TabelaVeículos}: stores schedule times of bus itinerary segments.
\end{itemize}

\subsection{Case Study - Bus Line 829} \label{sec:case_study}

The bus line 829 ``Universidade Positivo'' (``Alimentador'') was chosen for a case study. It is circular, i.e., the same start and end stops, with single-direction trips. In addition, it has few stops that make visualization and interpretation of the data easier.

\begin{figure}[H]
\begin{center}
\includegraphics[width=.70\columnwidth]{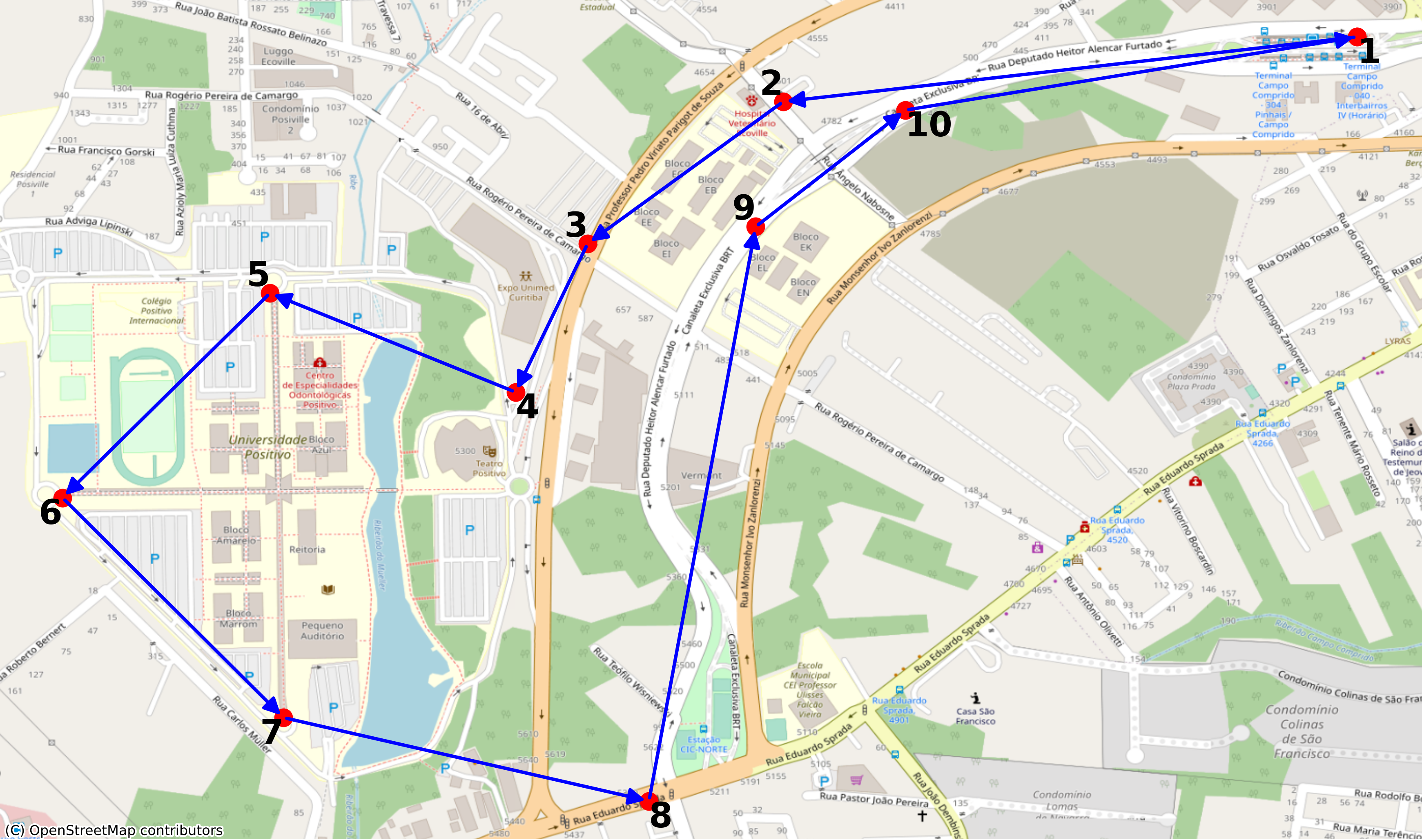}
\caption{Programmed itinerary of line 829 that starts at point 1, passes through intermediate points 2 to 10, and returns to the starting point.}
\label{fig:mapa_itinerario_829}
\end{center}
\end{figure}

\textbf{Table \ref{tab:itinerario_829}} shows the scheduled itinerary of bus line 829, containing bus stop names and sequences. The bus route starts at stop 1 in ``Terminal Campo Comprido'', reaches intermediate stops 2 to 10, and returns to the starting stop 1, as shown in \textbf{Figure \ref{fig:mapa_itinerario_829}}.

\begin{table}[h]
\centering
\footnotesize
\caption{Scheduled itinerary of bus line 829 with bus stops.}
\label{tab:itinerario_829}
\begin{tabular}{@{}ll@{}}
\hline\hline
\textbf{Bus stop} & \textbf{Seq.} \\
\hline
Terminal Campo Comprido & 1 \\
R. Angelo Nebosne, 75 & 2 \\
R. Prof. Pedro Viriato Parigot de Souza, 4716 & 3 \\
R. Prof. Pedro Viriato Parigot de Souza, 5136 & 4 \\
R. Casemiro Augusto Rodacki, 233 & 5 \\
R. Carlos Müller, 331 & 6 \\
R. Carlos Müller, 871 & 7 \\
R. Eduardo Sprada, 5273 & 8 \\
R. Dep. Heitor Alencar Furtado, 5181 & 9 \\
R. Dep. Heitor Alencar Furtado, 4900 & 10 \\
Terminal Campo Comprido & 1 \\
\hline\hline
\end{tabular}
\end{table}

The scenario was built using real logs from 07/11/2022 of bus BA020.
A round trip occurs between 06:04 to 06:32, during which there is no loss of GPS data. Therefore, this scenario is a suitable case to verify the application of the proposed algorithm and evaluate interpolation errors, simulating communication failures. Some logs are then deleted within specific time intervals. In this case, the map-matching algorithm does not detect the vehicle passing at some points, and the proposed algorithm can recover the information using interpolation. Points 3, 5, and 8 of bus line 829 were chosen to be removed from the original data set, according to \textbf{Table \ref{tab:parametros_teste}}.

\begin{table}[H]
\centering
\footnotesize
\caption{Case study parameters for line 829.}
\label{tab:parametros_teste}
\begin{tabular}{@{}ll@{}}
\hline\hline
\textbf{Line} & 829 - Universidade Positivo \\
\hline
\textbf{Bus} & BA020 \\
\textbf{Date} & 7/11/2022 \\
\textbf{Return time} & 06:04 to 06:32 \\
\hline
\multirow{3}{*}{\textbf{Failure Interval}}
& 06:15 to 06:16 at point 3 \\
& 06:17 to 06:19 at point 5 \\
& 06:26 to 06:28 at point 8 \\
\hline\hline
\end{tabular}
\end{table}

\textbf{Table \ref{tab:resultado_map_829}} presents the result of \textbf{steps 1}, \textbf{2} and part of \textbf{step 3}. The procedure marks the exact time when bus BA020 passes at stops (map matching), creates a temporal sequencing (\textbf{step 2}), locates the itinerary, and assigns a sequence number to each log (part of \textbf{step 3}).

\begin{table}[h]
\centering
\footnotesize
\caption{Results of applying steps 1, 2, and part of step 3 to the case study.}
\label{tab:resultado_map_829}
\begin{tabular}{lll}
\hline\hline
\textbf{Bus stop} & \textbf{Time} & \textbf{Sequence} \\
\hline
Terminal Campo Comprido & 06:04:51 & 1 \\
R. Dep. Heitor Alencar Furtado, 4900 & 06:14:08 & 10 \\
R. Angelo Nebosne, 75 & 06:14:36 & 2 \\
R. Prof. Pedro Viriato Parigot de Souza, 5136 & 06:16:43 & 4 \\
R. Carlos Müller, 331 & 06:19:30 & 6 \\
R. Carlos Müller, 871 & 06:21:06 & 7 \\
R. Dep. Heitor Alencar Furtado, 5181 & 06:28:30 & 9 \\
R. Dep. Heitor Alencar Furtado, 4900 & 06:29:06 & 10 \\
Terminal Campo Comprido & 06:31:41 & 1 \\
\hline\hline
\end{tabular}
\end{table}

\begin{figure}[H]
\begin{center}
\includegraphics[width=.70\columnwidth]{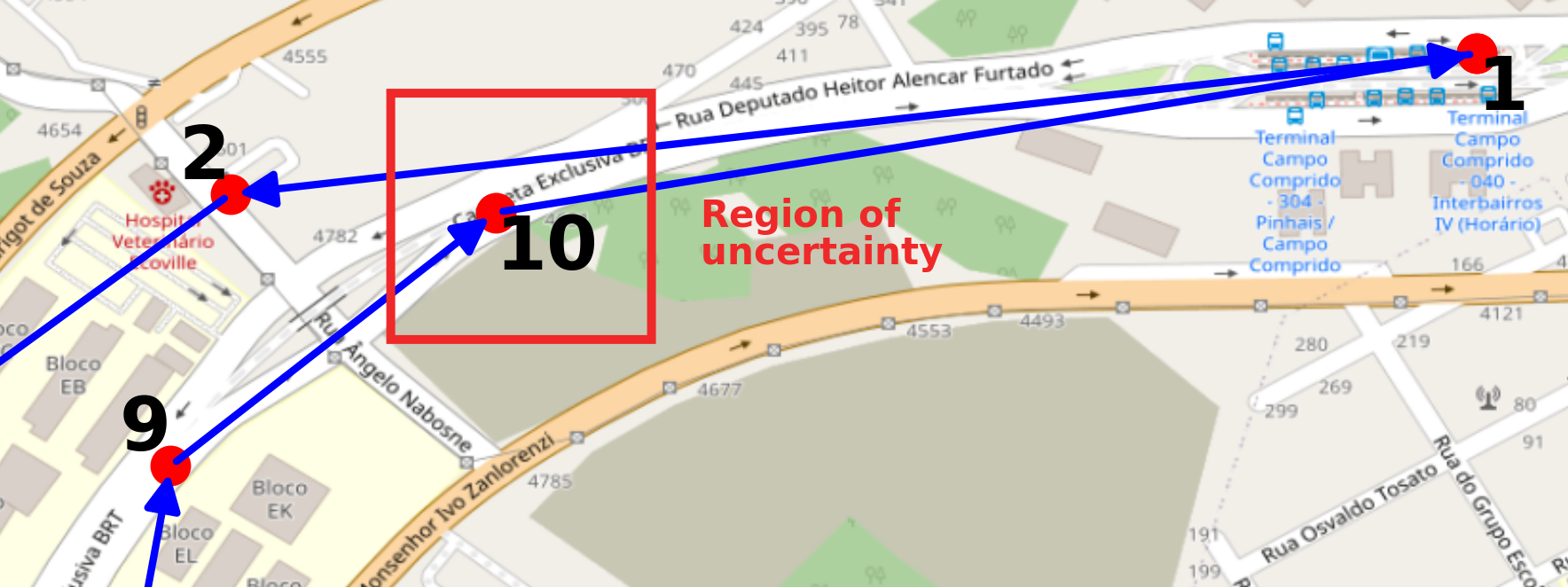}
\caption{Region of uncertainty in which the map matching algorithm generates a marking error.}
\label{fig:regiao_incerteza_829}
\end{center}
\end{figure}

However, \textbf{Table~\ref{tab:resultado_map_829}} contains some inconsistencies. For example, the stop ``R. Dep. Heitor Alencar Furtado, 4900'' marked at 06:14:08 is incorrect because it is at the end of bus itinerary. A close examination reveals that the bus route between points 1 and 2 passes very close to point 10, as illustrated in \textbf{Figure \ref{fig:regiao_incerteza_829}}. In this case, the map-matching algorithm generates a markup error. This algorithm is thus insufficient to handle logs properly. The marking error is identified by combining the result of map-matching with the proposed \textbf{Algorithm~\ref{alg:det_iti}}.

The proposed algorithm identifies gaps in the sequence after completing \textbf{step 3}. Due to the communication failure simulation, it identifies points $3$, $5$, and $8$ are lacking as shown in \textbf{Table \ref{tab:resultado_final_829}}.
The incorrect marking was deleted, and points 3, 5, and 8 were added to the itinerary due to temporal interpolation. This itinerary corresponds to the complete sequence of bus stops registered for line 829 with temporal information.

\begin{table}[h]
\centering
\footnotesize
\caption{Final result of applying Algorithm~\ref{alg:det_iti} to the case study.}
\label{tab:resultado_final_829}
\begin{tabular}{@{}lll@{}}
\hline\hline
\textbf{Bus stop} & \textbf{Time} & \textbf{Sequence} \\
\hline
Terminal Campo Comprido & 06:04:51 & 1 \\
R. Angelo Nebosne, 75 & 06:14:36 & 2 \\
R. Prof. Pedro Viriato Parigot de Souza, 4716 & 06:15:39 & 3 \\
R. Prof. Pedro Viriato Parigot de Souza, 5136 & 06:16:43 & 4 \\
R. Casemiro Augusto Rodacki, 233 & 06:18:07 & 5 \\
R. Carlos Müller, 331 & 06:19:30 & 6 \\
R. Carlos Müller, 871 & 06:21:06 & 7 \\
R. Eduardo Sprada, 5273 & 06:24:48 & 8 \\
R. Dep. Heitor Alencar Furtado, 5181 & 06:28:30 & 9 \\
R. Dep. Heitor Alencar Furtado, 4900 & 06:29:06 & 10 \\
Terminal Campo Comprido & 06:31:41 & 1 \\
\hline\hline
\end{tabular}
\end{table}

In evaluating the interpolation error, data from the movement of buses on line 829 during a whole day from 06:04 to 23:19 were used. Known points were randomly taken from the original data set, simulating communication failures. The estimation error $err_i=|t_i-\hat{t}_i|$ was calculated from the known real values $(p_i,t_i)$ of the points removed and the estimated values $(p_i,\hat{t}_i )$.

\begin{figure}[H]
\begin{center}
\includegraphics[width=0.5\columnwidth]{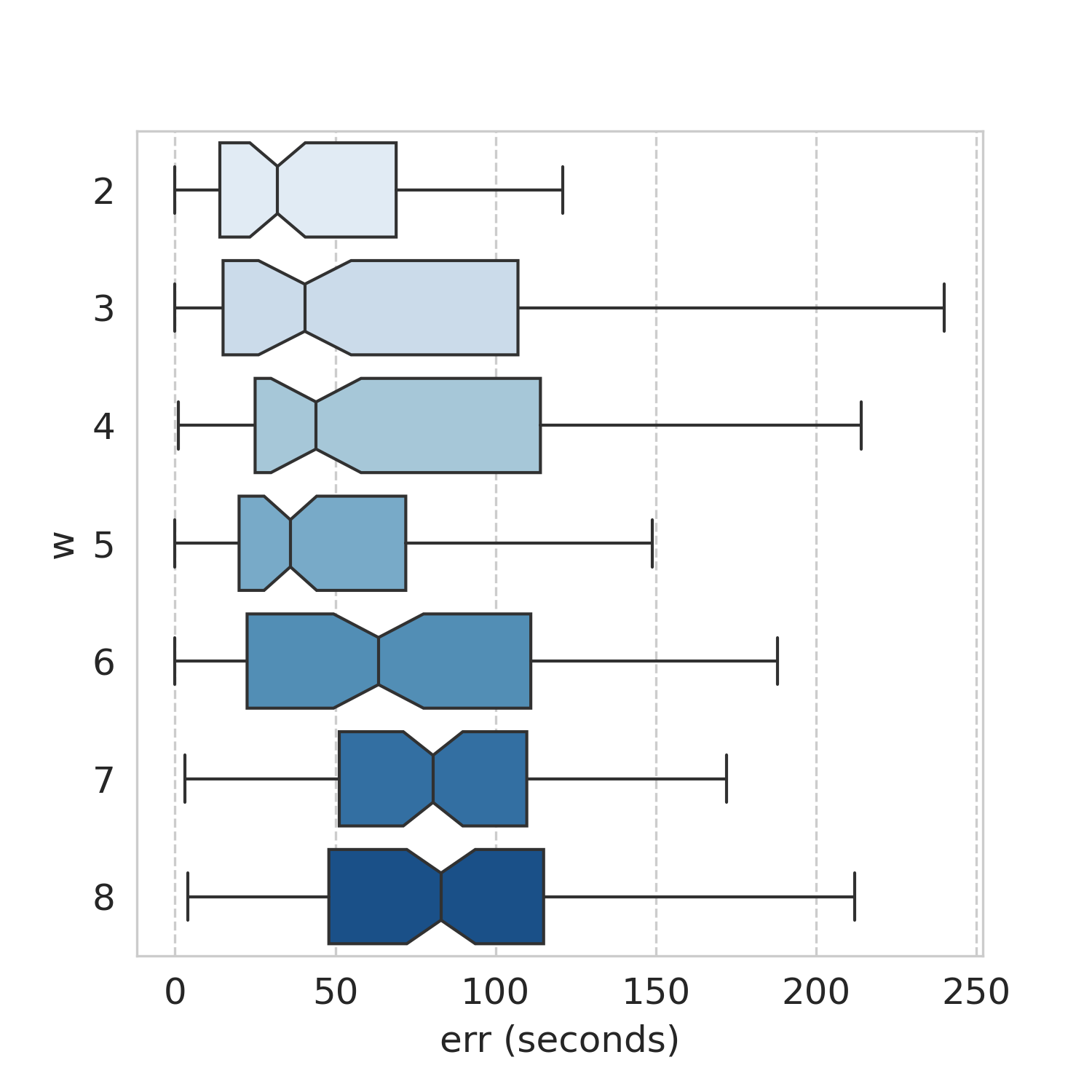}
\caption{Interpolation error in seconds on line 829 for different values of $w$ (a missing midpoint corresponds to $w=2$).}
\label{fig:boxplot_err}
\end{center}
\end{figure}

Error measures are computed as a function of the number of consecutive bus stops missing. In this experiment, error measures are calculated in seconds for 1 to 7 consecutive missing points, or $w \in \{2, 3, \dots, 8\}$. For each case, $100$ samples without replacement were used to generate the result in \textbf{Figure \ref{fig:boxplot_err}}. It is observed that the interpolation error increases with the number of missing points. For most cases, the error ranges from less than 1 min to approximately 2 min (125 seconds).

This result suggests that the uncertainty introduced by interpolation is acceptable. A delay or advance of 2 minutes can be considered tolerable in an urban bus transport system.
However, a closer look is needed to understand better which lines are most affected by the interpolation error at which times of day.

\subsection{Database Assessment}\label{sec:limitacoes}

The proposed algorithm was applied to logs on 07/11/2022 to evaluate the ability to detect itineraries using the entire database. The result is compared with the algorithm of \cite{Peixoto2020}, which uses bus schedule tables. \textbf{Table \ref{tab:horarios_x_algoritmo}} shows the total number of assignments each algorithm makes to a valid itinerary by type of bus line.

\begin{table}[h]
\centering
\footnotesize
\caption{Comparison between the number of tags assigned to a valid itinerary according to~\citep{Peixoto2020} and the proposed algorithm by type of bus line.
The percentage values are relative to the number of matches made by the map-matching algorithm.}
\label{tab:horarios_x_algoritmo}
\begin{tabular}{@{}l|cc|cc@{}}
\hline\hline
\multirow{2}{*}{\textbf{Line type}}
& \multicolumn{2}{c}{\textbf{\citep{Peixoto2020}}} & \multicolumn{2}{|c}{\textbf{Proposed}}\\
\cline{2-5}
&\textbf{Tags} &\textbf{\%} &\textbf{Tags} &\textbf{\%}\\
\hline
ALIMENTADOR & 160,750 & 70.62\% & 225,434 & 99.03\% \\
CONVENCIONAL & 52,338 & 62.02\% & 84,310 & 99.90\% \\
EXPRESSO & 21,986 & 62.41\% & 35,206 & 99.94\% \\
JARDINEIRA & 234 & 46.89\% & 499 & 100.00\% \\
LIGEIRÃO & 2,988 & 63.86\% & 4,677 & 99.96\% \\
LINHA DIRETA & 8,421 & 70.37\% & 11,879 & 99.26\%\\
MADRUGUEIRO & 5,659 & 98.26\% & 5,455 & 94.72\% \\
TRONCAL & 26,136 & 75.84\% & 34,447 & 99.96\% \\
\hline
\textbf{TOTAL}
& \textbf{278,512} & \textbf{68.83\%} & \textbf{401,907} & \textbf{99.33\%} \\
\hline\hline
\end{tabular}
\end{table}

It is observed that the proposed algorithm provides a global increase from $68.83\%$ to $99.33\%$ in itinerary traceability gain. The new algorithm presents a result $44.31\%$ better than ~\citep{Peixoto2020}. Except for lines of MADRUGUEIRO that should be further investigated, all lines of other types benefit. This result increases valid data in the database, preventing data from being discarded due to not being associated with any itinerary. 

\begin{table}[h]
\centering
\footnotesize
\caption{Distribution of the tags of \textbf{Table~\ref{tab:horarios_x_algoritmo}} that had two error types: i) out of order and ii) missing bus stops for the proposed algorithm.}
\label{tab:resultado_parcial_todos}
\begin{tabular}{@{}l|cccc@{}}
\hline\hline
\multirow{2}{*}{\textbf{Line type}} &\multicolumn{4}{c}{\textbf{Tags with errors}} \\
\cline{2-5}
& \textbf{i} & \textbf{ii} & \textbf{Total} & \textbf{\%} \\
\hline
ALIMENTADOR & 15,764 & 21,176 & 36,940 & 16.39\% \\
CONVENCIONAL & 2,432 & 7,068 & 9,500 & 11.27\% \\
EXPRESSO & 487 & 2,139 & 2,626 & 7.46\% \\
JARDINEIRA & 0 & 12 & 12 & 2.40\% \\
LIGEIRÃO & 83 & 193 & 276 & 5.90\% \\
LINHA DIRETA & 126 & 162 & 288 & 2.42\% \\
MADRUGUEIRO & 1,470 & 283 & 1,753 & 32.14\% \\
TRONCAL & 478 & 1,896 & 2,374 & 6.89\% \\
\hline
\textbf{TOTAL}
& \textbf{20,840} & \textbf{32,929} & \textbf{53,769} & \textbf{13.38\%} \\
\hline\hline
\end{tabular}
\end{table}

Although the tags of \textbf{Table ~\ref{tab:horarios_x_algoritmo}} are assigned to valid itineraries at the end of the proposed algorithm, some of them had errors in the sequence of stops or missing bus stops. 
\textbf{Table \ref{tab:resultado_parcial_todos}} shows 
the distribution of tags that had two error types: i) out of order and ii) missing bus stops for the proposed algorithm.
The percentage of $13.38\%$ of tags is practically all corrected after the adjustment of sequence and missing points performed at the end of the proposed algorithm.

\begin{figure}[H]
\begin{center}
\includegraphics[width=.60\columnwidth]{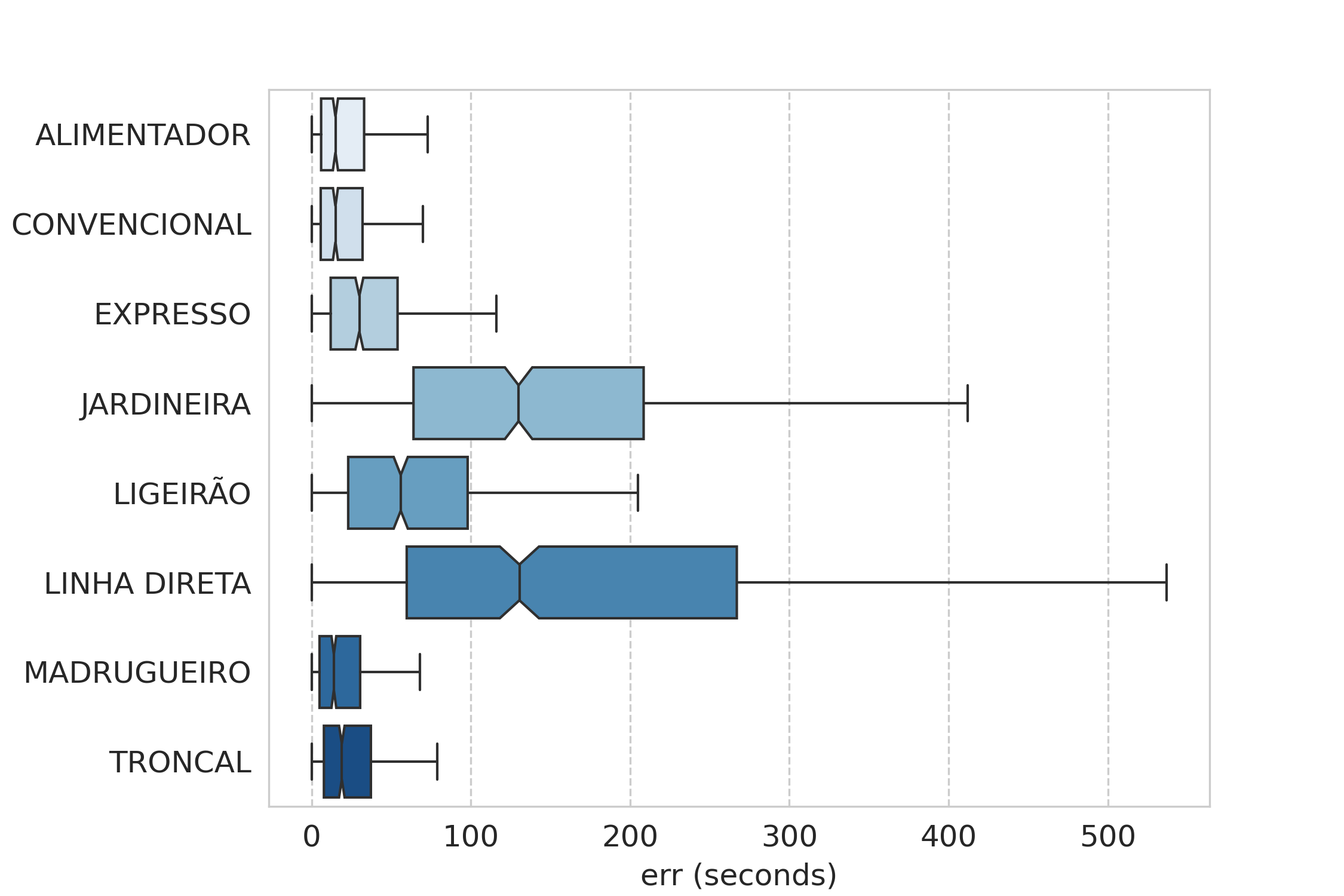}
\caption{Line type interpolation error.}
\label{fig:boxplot_err_todos}
\end{center}
\end{figure}

The interpolation error by line type is evaluated similarly to 
\textbf{Section~\ref{sec:case_study}} but using all bus lines in the database that take only complete paths with real values. \textbf{Figure \ref{fig:boxplot_err_todos}} shows the interpolation error in seconds by line type. Bus lines ALIMENTADOR, CONVENCIONAL, EXPRESSO, MADRUGUEIRO, and TRONCAL have an error between 0 and 1 min approximately. On the other hand, JARDINEIRA, LIGEIRÃO, and LINHA DIRETA are more susceptible to interpolation errors. The big errors for JARDINEIRA and LINHA DIRETA might be due to the long distances between bus stops of these lines, whose time interpolation can be significantly affected by road traffic conditions.

\subsection{Temporal Assessment of Bus Service}\label{sec:temporal}

A bus service is provided according to the timetables of bus lines. Therefore, users can estimate the time interval between consecutive buses of a given bus line. However, from a user's perspective, at a single bus stop, several buses from different bus lines interact to provide the bus service to that stop. In this case, how often do buses serve a particular bus stop (eventually by different bus lines)? This section provides a temporal assessment of bus services by identifying well-served urban areas with many buses and bus lines. It considers the database from 07/11/2022 to 07/15/2022. The GPS bus trajectories are tracked according to the method described in \textbf{Section~\ref{sec:algoritmo}}.

The bus service is evaluated in a time window of 10 minutes, representing an expected waiting time for most users. The number of buses that pass a stop is counted considering consecutive time windows of $10$ minutes. This way, buses within a time window of $10$ minutes are counted, a shift of 1 minute is then given to the window, allowing buses to be counted within the next 10 minutes, and so on. For each bus stop, a time series represents the number of buses observed in a 10-minute time interval in each minute of the day.

\begin{figure}[h]
\centering
\begin{minipage}{.45\columnwidth}
    \centering
    \includegraphics[width=1\columnwidth]{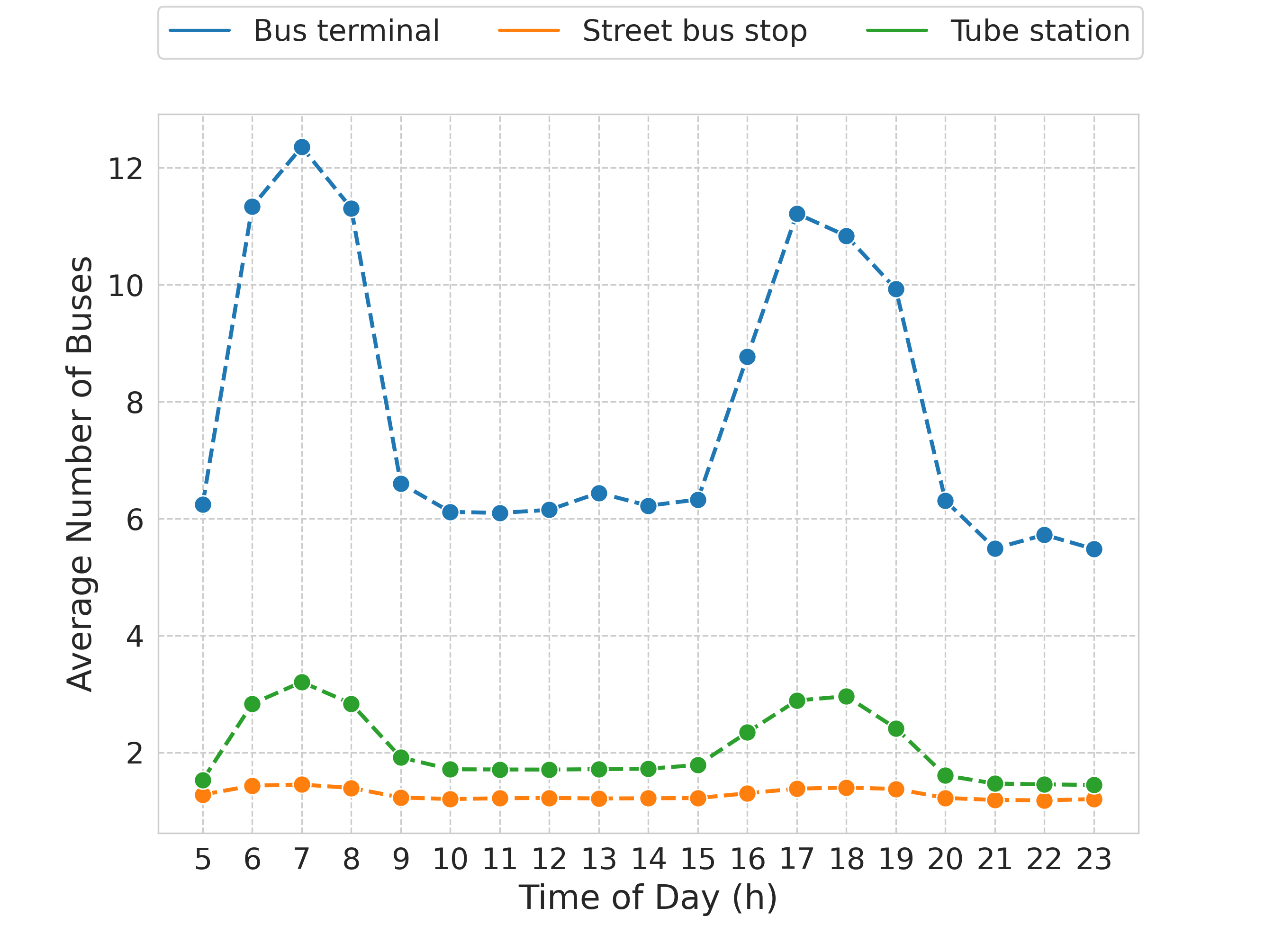}
    \caption{Average number of buses observed at a stop of terminals, street stops, and tube stations in a 10-minute moving window from 5:00 to 23:00.}
    \label{fig:bus_availability}
\end{minipage}%
\hfill 
\begin{minipage}{.45\columnwidth}
    \centering
    \includegraphics[width=1\columnwidth]{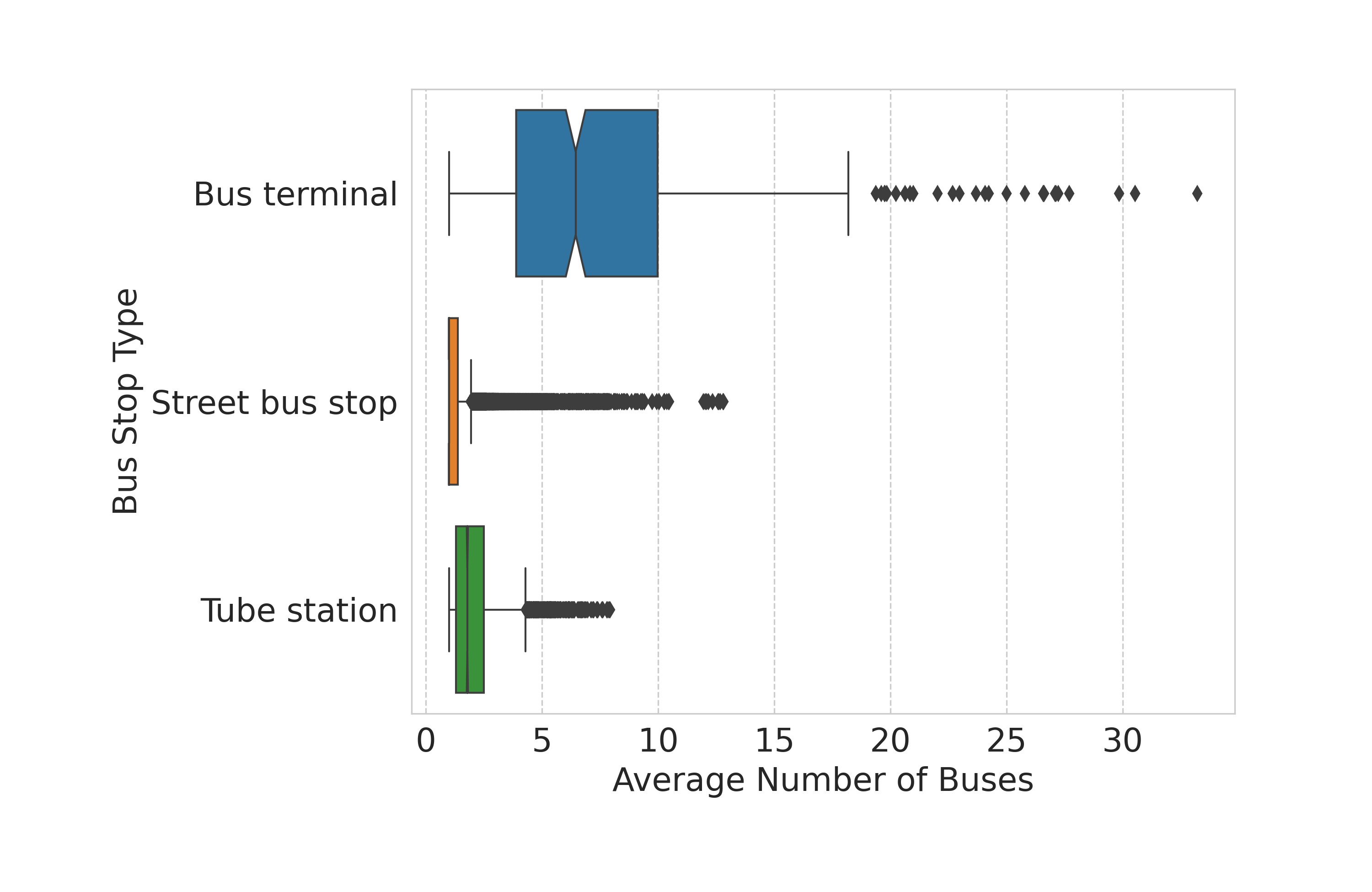}
    \caption{Boxplots showing the average number of buses observed at a stop of terminals, street stops, and tube stations in a 10-minute moving window.}
    \label{fig:daily_avg_bus_availability}
\end{minipage}
\end{figure}

\textbf{Figure \ref{fig:bus_availability}} shows the average number of buses observed at a bus stop within a time window of 10 min shifted from 5:00 to 23:00 and aggregated into three categories of bus stops (terminal, street stop, and tube station). All bus stops inside a terminal are considered a single stop for capturing the number of buses available at a given time. \textbf{Figure \ref{fig:bus_availability}} shows a peak from 6:00 to 8:00, with a maximum of around 7:00, and from 16:00 to 19:00, with a maximum of around 17:00. This behavior is relevant for terminals and tube stations with minor effects on street stops. Moreover, terminals present up to $4$ times more buses than tube stations at peak hours, as expected, because they are hubs integrating different bus lines. Terminals and tube stations play an important role in Curitiba's transport system because they offer users the possibility of transferring with a single fare.


The time series of \textbf{Figure~\ref{fig:bus_availability}} are then aggregated in a day, obtaining the average number of buses at a stop for terminals, street stops, and tube stations as shown in the boxplots of \textbf{Figure~\ref{fig:daily_avg_bus_availability}}. Terminals have the highest average number of buses, as expected. Street stops and tube stations have fewer buses but stand out as outliers, ranging from $2$ to $13$ buses on average. This means that street bus stops can also act as hubs if some integration between bus lines could be provided.

The outliers of \textbf{Figure~\ref{fig:daily_avg_bus_availability}} represent an opportunity to improve the bus service because they have bus stops with a high frequency of buses. If the stops are close enough to each other, a hub can be built to allow connections between the respective bus lines. For instance, if temporal integration (with payment of a single fare) is allowed in certain regions of interest, new links between bus lines can be made in the network, eventually shortening distances and trip times. The regions of interest contain stops with a high concentration of buses, as shown in \textbf{Figure \ref{fig:regioes_interesse}}.


It is a heat map obtained from outliers of street stops and tube stations (terminals are far from each other and usually do not have the potential for integration). The map highlights red regions with a high density of stops and frequency of buses. The most dense regions partially follow the North/South transportation corridor. It means that buses running in this corridor have great potential to improve bus line connections in areas other than terminals. Based on this result, we aim to build bus stop clusters to behave like virtual terminals.

\subsection {Virtual Terminal Evaluation}

\begin{figure}[h]
\centering
\begin{minipage}{.45\columnwidth}
    \centering
    \includegraphics[width=0.8\columnwidth]{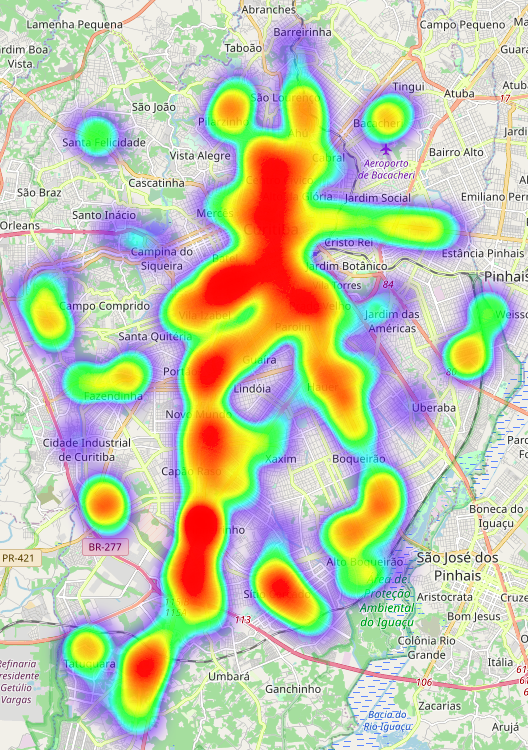}
    \caption{Heat map of regions with a high density of stops and frequency of buses obtained from outliers of street stops and tube stations.}
    \label{fig:regioes_interesse}    
\end{minipage}%
\hfill 
\begin{minipage}{.45\columnwidth}
    \centering
    \includegraphics[width=0.8\columnwidth]{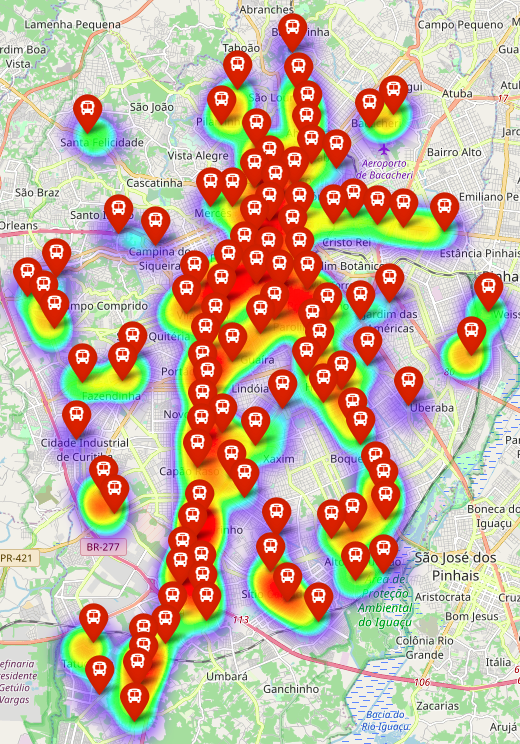}
    \caption{Centroids of $104$ clusters.}
    \label{fig:centroides_clusteres}
\end{minipage}
\end{figure}

\begin{figure}[h]
\centering
\begin{minipage}{.45\columnwidth}
    \centering
    \includegraphics[width=1\columnwidth]{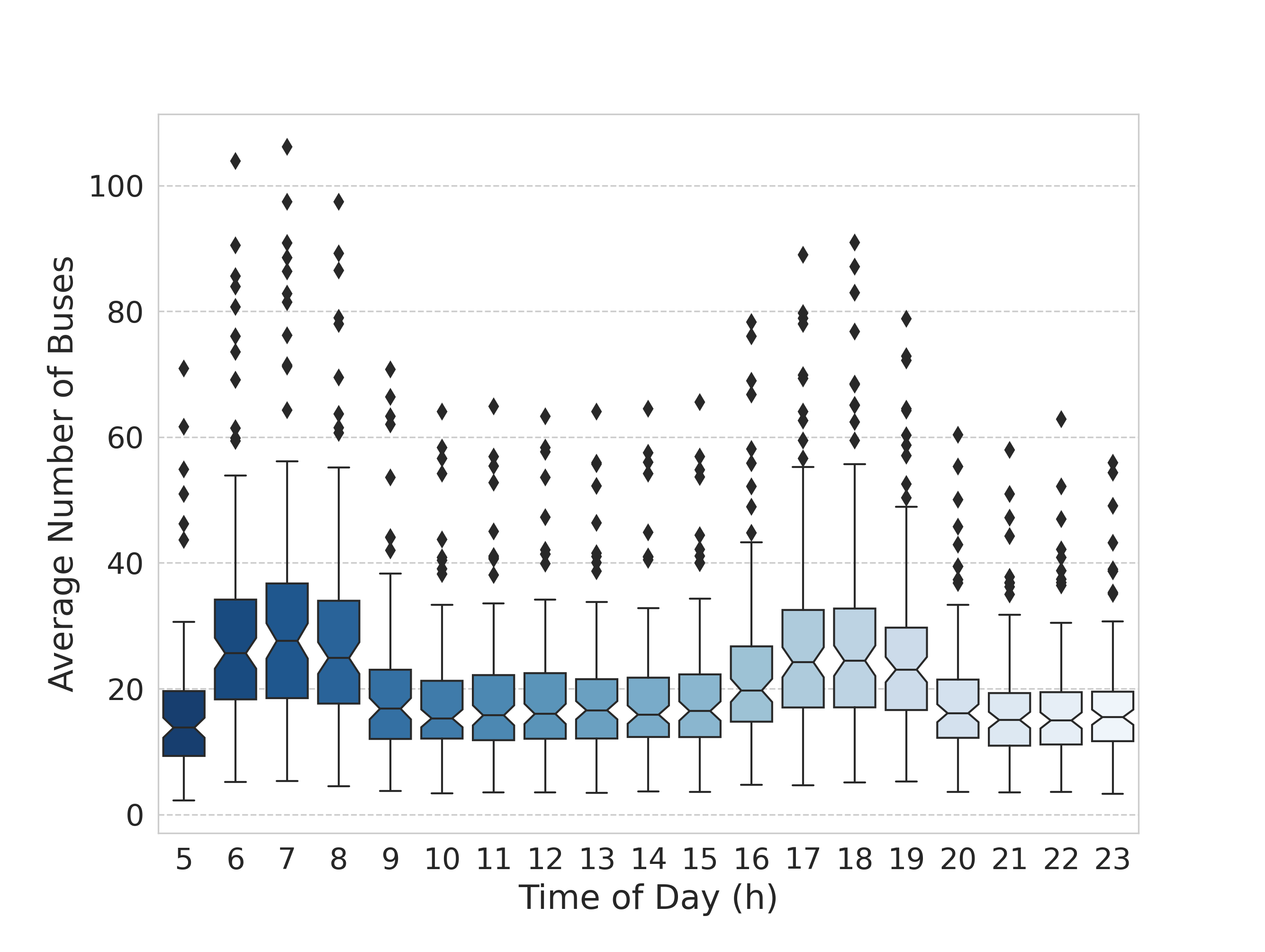}
    \caption{Boxplots showing the average number of buses observed at a cluster in a 10-minute moving window from 5:00 to 23:00.}
    \label{fig:cluster_avg_bus_availability}   
\end{minipage}%
\hfill 
\begin{minipage}{.45\columnwidth}
    \centering
    \includegraphics[width=1\columnwidth]{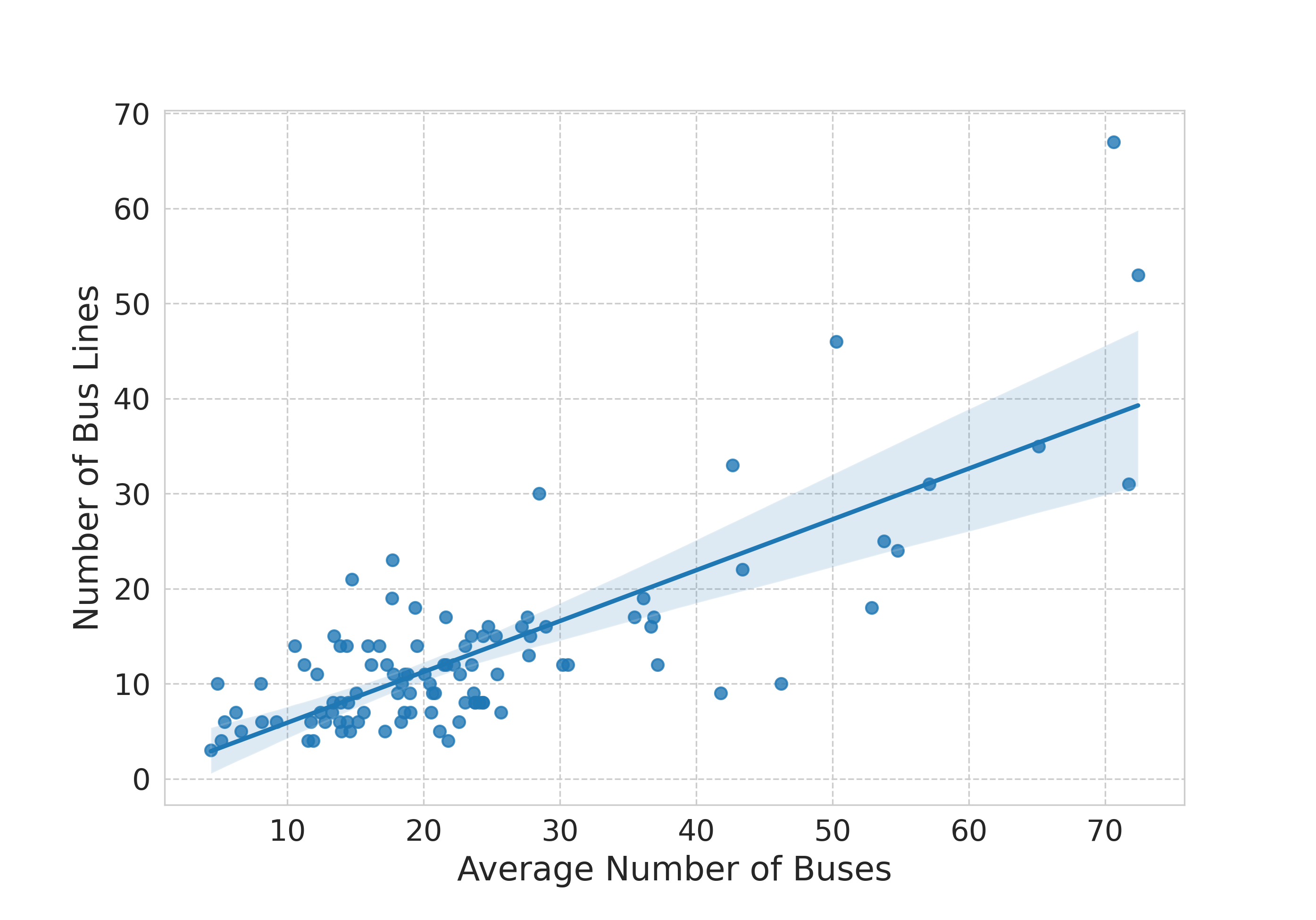}
    \caption{Scatter plot of the average number of buses observed in a 10-minute moving window and the number of bus lines of all 104 clusters.}
    \label{fig:corr_bus_aval_lines}
\end{minipage}
\end{figure}

The results of \textbf{Algorithm~\ref{alg:cluster}} are shown in \textbf{Figure~\ref{fig:centroides_clusteres}}. It shows centroids of 104 clusters computed with 27 bus stops, each serving 15 bus lines on average and covering 2,472 bus stops. \textbf{Figure \ref{fig:cluster_avg_bus_availability}} shows the average number of buses observed at a cluster in a 10-minute moving window from 5:00 to 23:00. Peak hours occur in the morning between 06:00 and 8:00 and between 16:00 and 19:00 in the afternoon. More than 100 buses, on average, are observed between 6:00 and 7:00.

There is a correlation between the average number of buses and the number of bus lines in a cluster according to \textbf{Figure \ref{fig:corr_bus_aval_lines}}. It shows a Pearson's correlation coefficient of $0.78$ with p-value < $0.001$. In other words, not only do many buses attend a cluster, but also many bus lines.

However, it is necessary to show that some ``synchronization'' exists between buses passing at cluster stops during the day. It is accomplished by evaluating the correlation between the bus time series of two stops of the same cluster.

\begin{figure*}[ht!]
\centering
\quad 
\subfloat[Morning]{
  \centering 
  \includegraphics[scale=0.25]{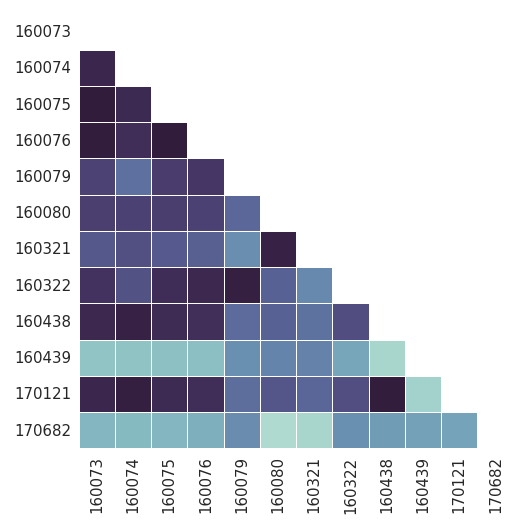}
  \label{fig:corr-170121-Morning}
}
\quad 
\subfloat[Midday]{
  \centering
  \includegraphics[scale=0.25]{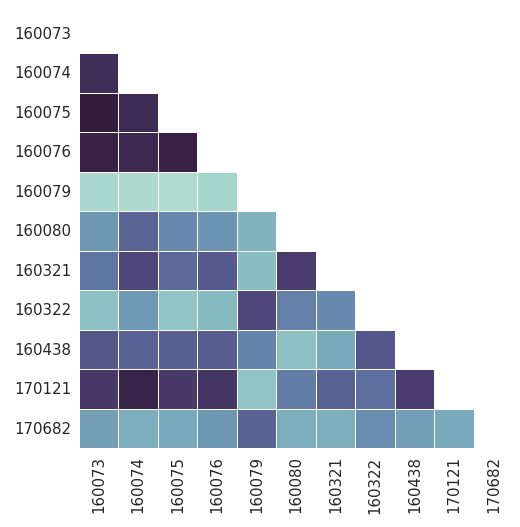}
  \label{fig:corr-170121-Afternoon}
}
\quad 
\subfloat[Evening]{
  \centering
  \includegraphics[scale=0.25]{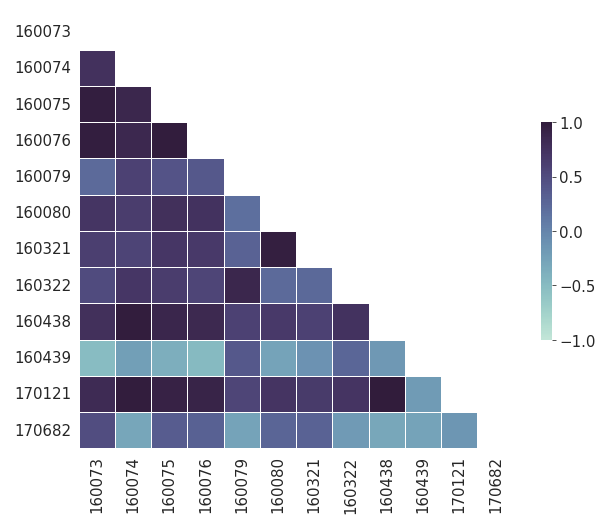}
  \label{fig:corr-170121-Evening}
}
\caption{Correlation matrix between bus stops in the cluster with centroid 170121 and all its neighbors for morning, midday, and evening.}
\label{fig:mapas}
\end{figure*}

Pearson's correlation between the time series of two bus stops is computed for all pairs of stops in a cluster. For instance, the correlation matrix between bus stops of the cluster with centroid 170121 is shown in \textbf{Figure~\ref{fig:mapas}}. 

\begin{figure}[H]
\centering
\begin{minipage}{.45\columnwidth}
    \centering
    \includegraphics[width=0.7\columnwidth]{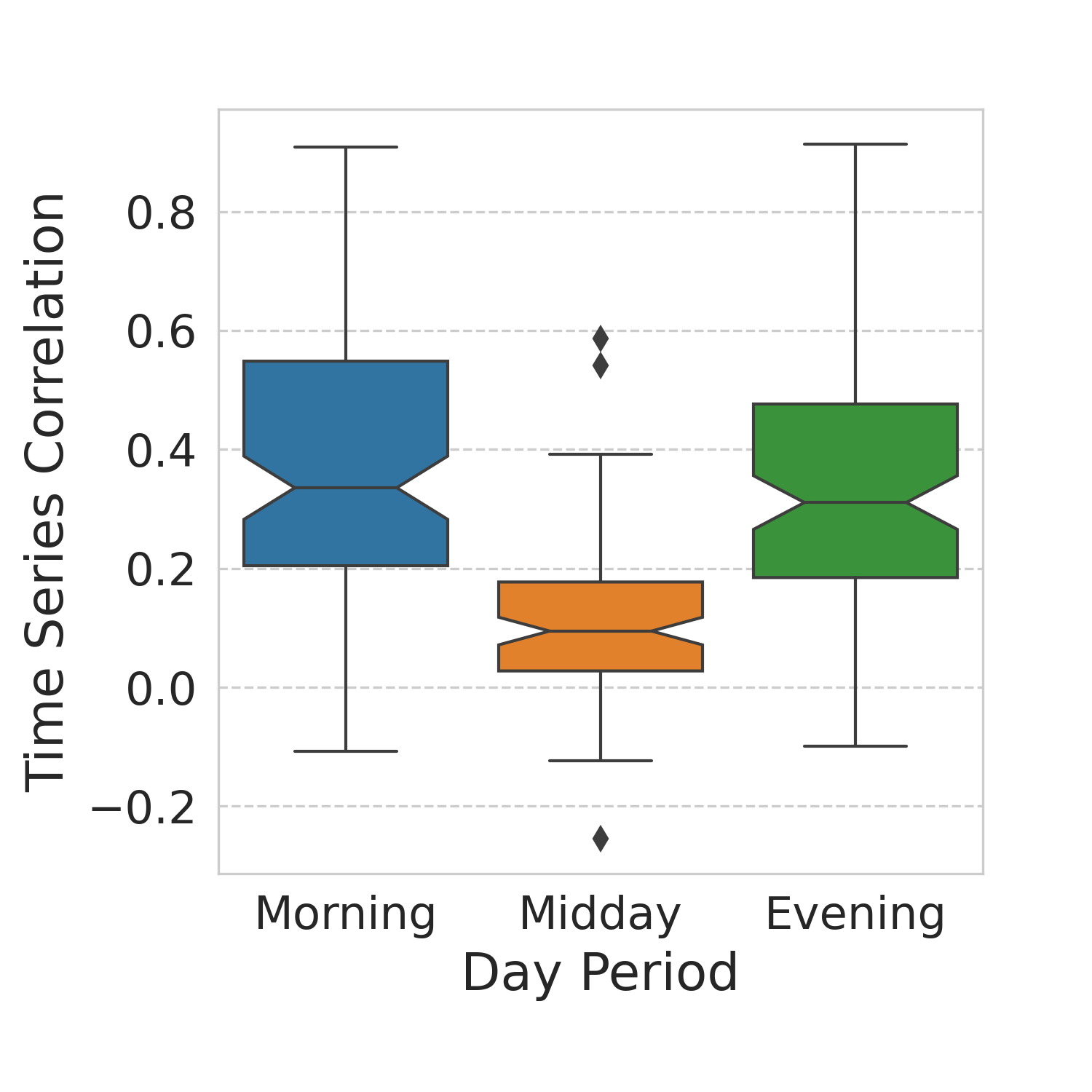}
    \caption{Correlation between time series of two bus stops averaged on all stop pairs of a cluster and all 104 clusters for morning, midday, and evening periods.}
    \label{fig:cluster_bus_stop_correlation}
\end{minipage}%
\hfill 
\begin{minipage}{.45\columnwidth}
    \centering
    \includegraphics[width=1\columnwidth]{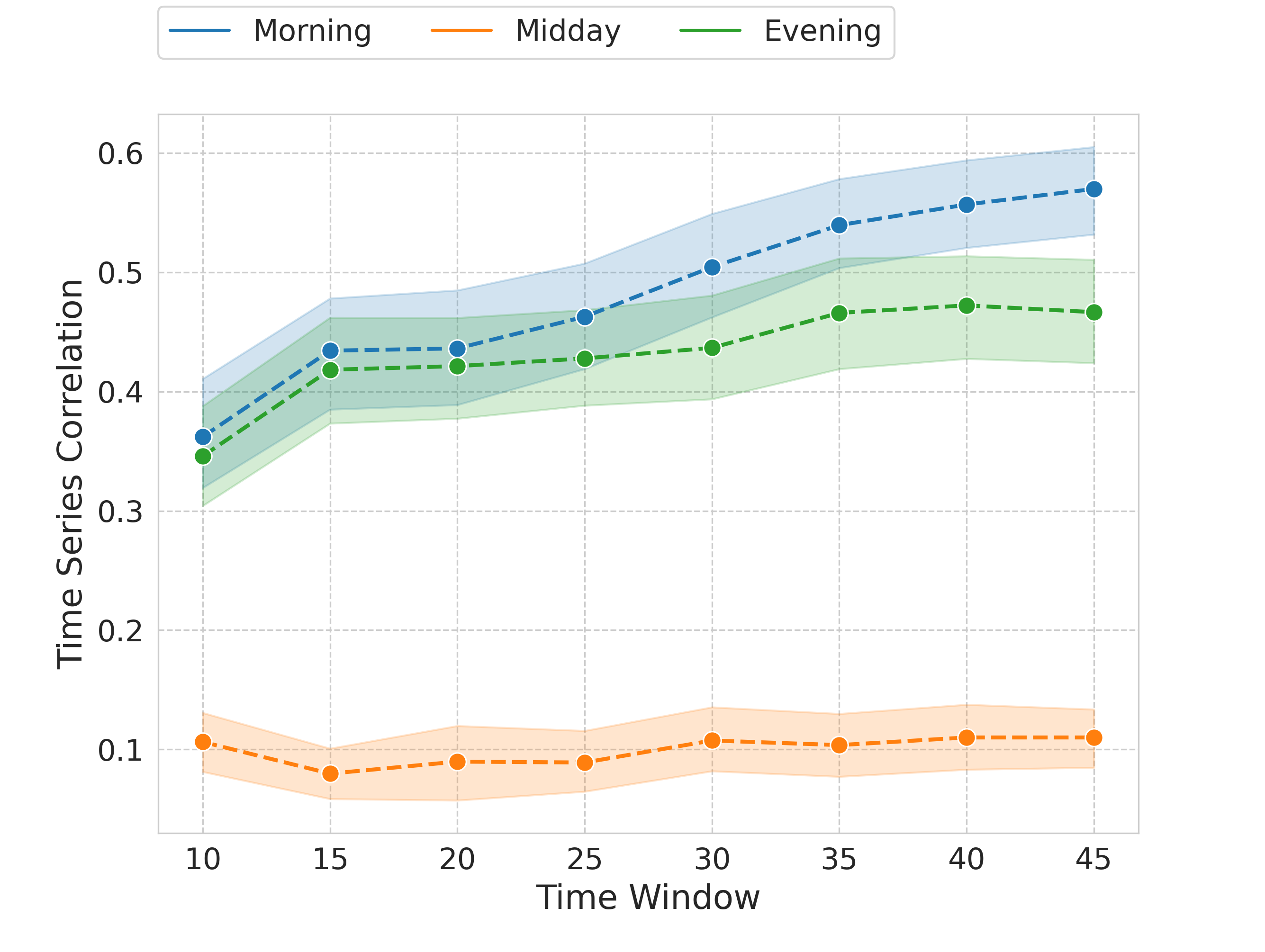}
    \caption{Observed correlation between time series for $\{10,15,20,25,30,35,40,45\}$-minute moving window.}
    \label{fig:correlation_time_window}
\end{minipage}
\end{figure}

A matrix is shown for each period of the day: i) morning from 6:00 to 9:00; ii) midday from 11:00 to 14:00. According to \textbf{Figure~\ref{fig:corr-170121-Morning}}, there are pairs of bus stops whose correlation achieves 0.75, which means that buses can meet each other more often in the morning. A similar behavior is observed in the evening from 17:00 to 20:00, as shown in \textbf{Figure~\ref{fig:corr-170121-Evening}}.
However, this behavior is not observed midday, according to \textbf{Figure~\ref{fig:corr-170121-Afternoon}}. Some pairs of bus stops with a strong correlation in the morning now show a weak correlation in the midday. The correlation between the time series of two bus stops is averaged on all bus stop pairs of a cluster and then averaged on all 104 clusters for morning, midday, and evening periods. The results are shown in \textbf{Figure~\ref{fig:cluster_bus_stop_correlation}}.

It can be seen that the average correlation is more relevant in the morning and evening. It means that buses passing in a cluster are better ``synchronized'' in the morning or evening on average, i.e., they can meet each other approximately simultaneously considering a time window of 10 min. It is then expected that users at bus stops of the same cluster can change between bus lines within 10 min on average.

When the time window increases, the correlation between the time series tends to increase; in other words, if the passenger is willing to wait longer, they will be more likely to make a bus line transition within the cluster. However, this is not valid for all periods of the day. \textbf{Figure \ref{fig:correlation_time_window}} shows the results obtained using time windows of $\{10,15,20,25,30,35,40,45\}$ minutes. The result suggests an improvement in ``synchronization'' during the morning and evening, but this does not happen at midday. The explanation is that during this period, as it is not a peak demand, transport companies remove several buses from the streets, negatively impacting the correlation between the time series.


\begin{figure*}[h!]
\centering
\quad 
\subfloat[Bus stops without clusters]{
  \centering 
  \includegraphics[scale=0.08]{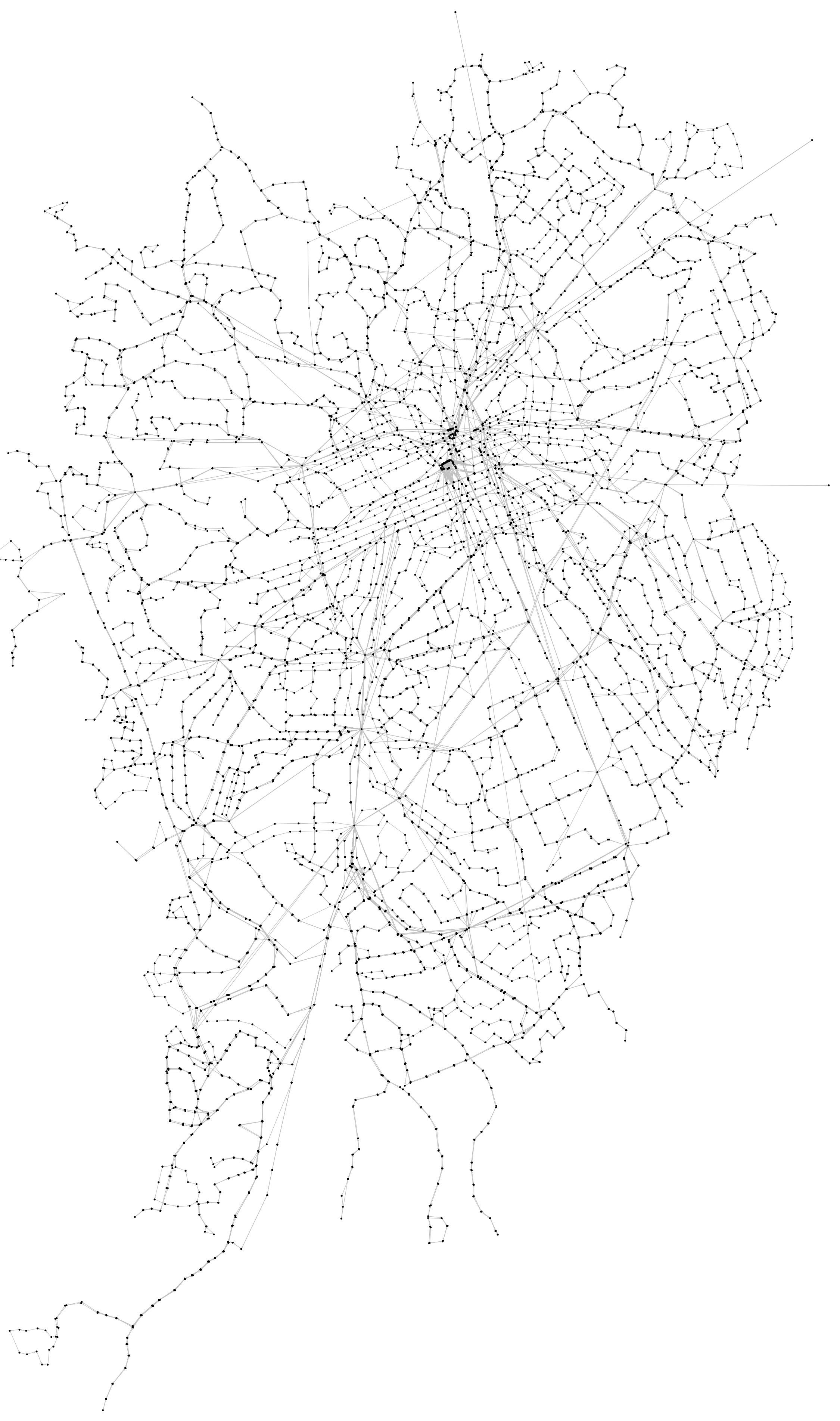}
  \label{fig:curitiba_sem_cluster}
}
\quad 
\subfloat[Bus stops with clusters]{
  \centering
  \includegraphics[scale=0.08]{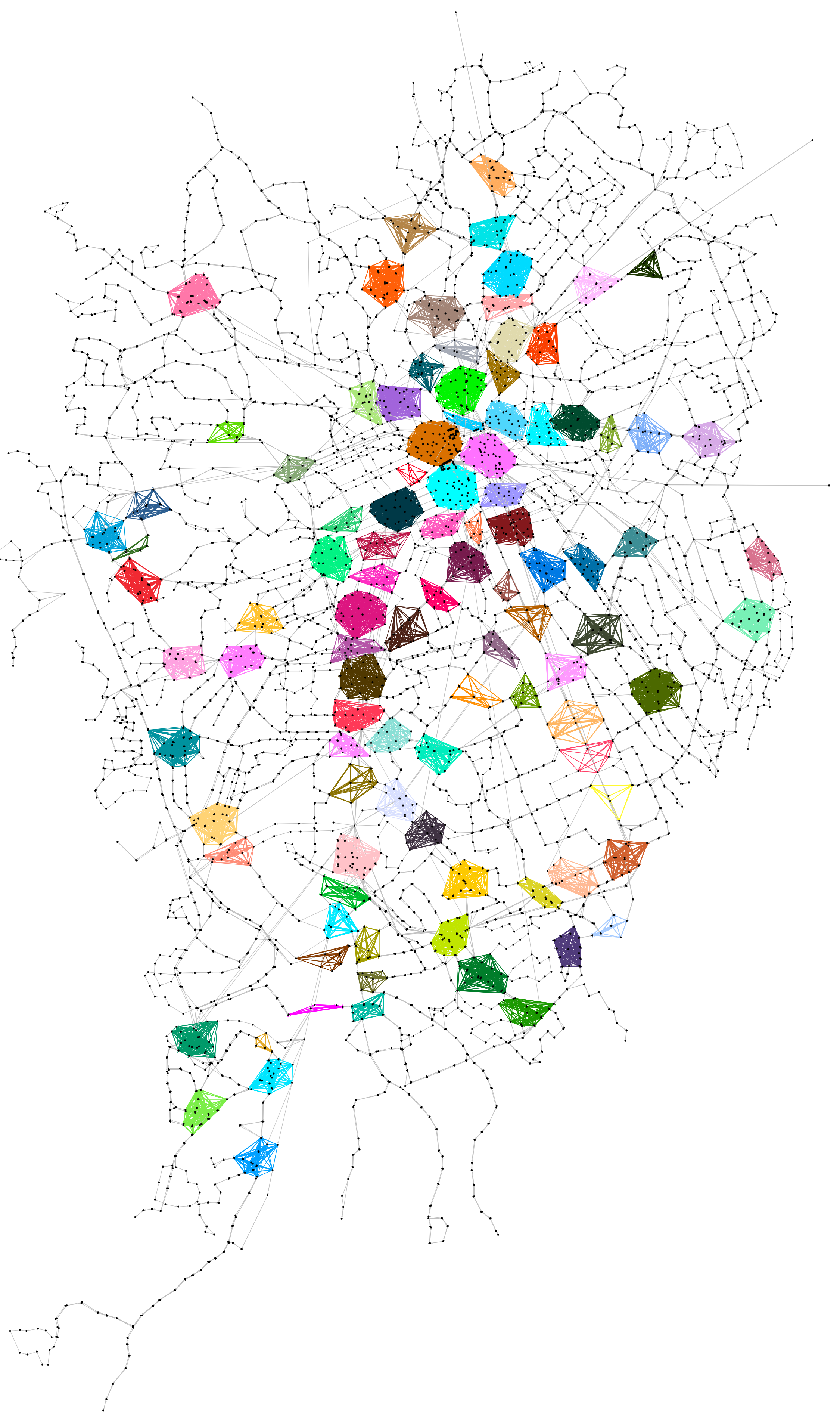}
  \label{fig:curitiba_com_cluster}
}
\caption{Curitiba bus transit network with and without clusters of bus stops.}
\label{fig:comparativo_mapas}
\end{figure*}

\subsection{Impact of Spatiotemporal Integration on Bus Transit}\label{sec:impact}

Given an origin and a destination in the bus transit network, the impact of spatiotemporal integration is measured by the distance traveled and the number of transfers made in a trip with and without clusters of bus stops. Each cluster can be seen as a virtual terminal where transfers are made between bus lines with a single fare. The question is how these additional possible connections between bus lines benefit a trip. Networks with and without clusters of bus stops are shown in \textbf{Figure~\ref{fig:comparativo_mapas}}.

\begin{figure}[h]
\centering
\quad 
\subfloat[Traveled distance (m)]{
  \centering 
  \includegraphics[scale=0.6]{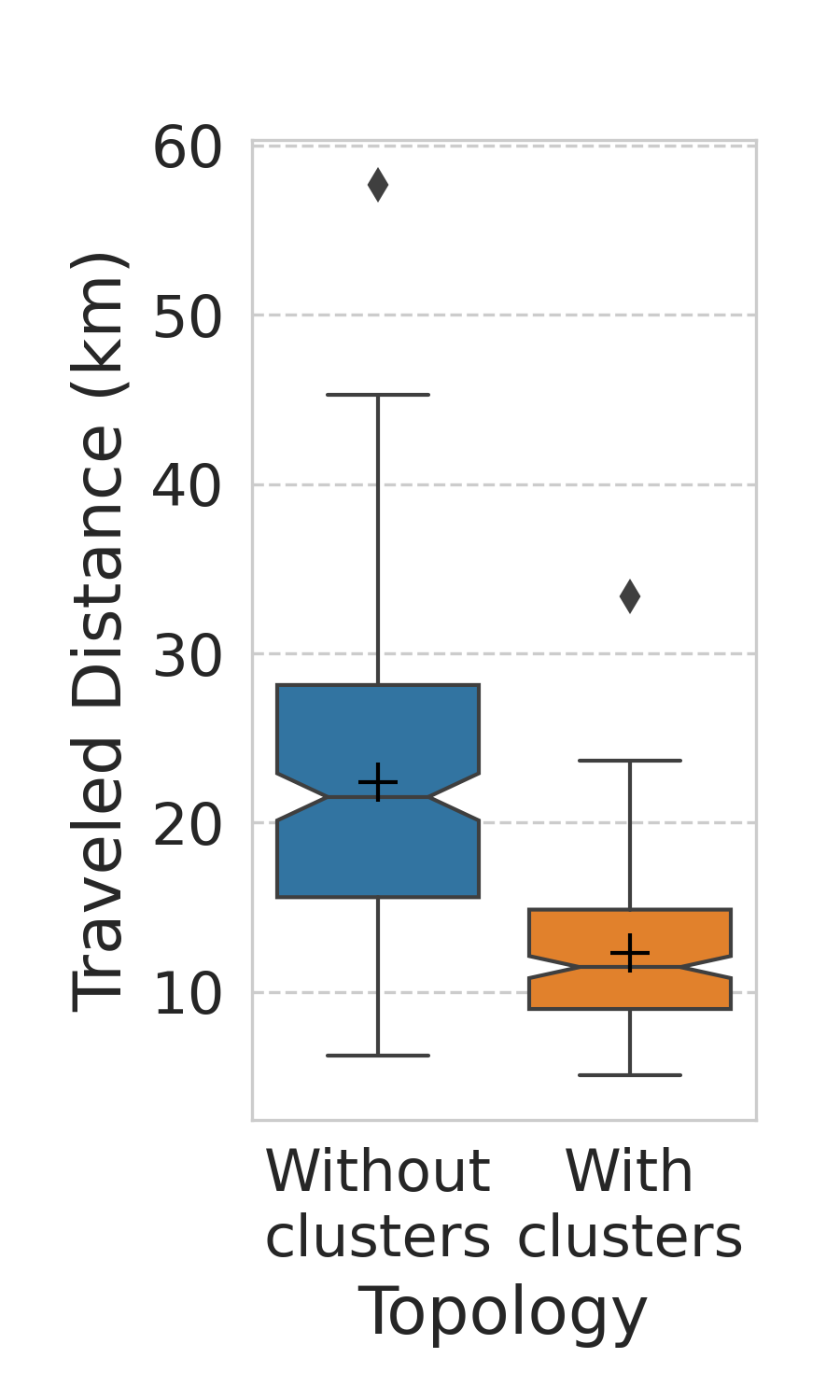}
  \label{fig:comparativo_custo}
}
\quad 
\subfloat[Number of transfers between bus lines]{
  \centering
  \includegraphics[scale=0.6]{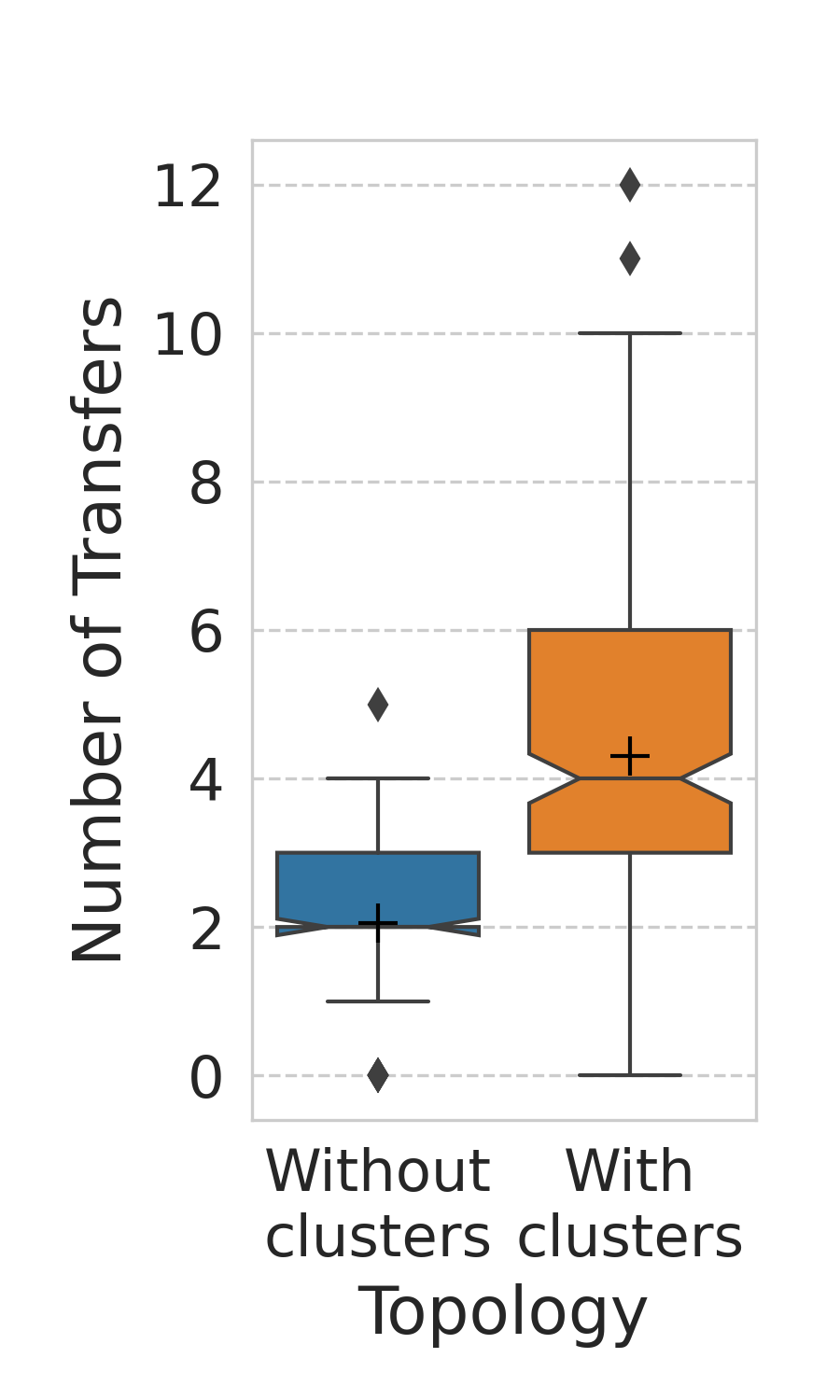}
  \label{fig:comparativo_custo_baldeacao}
}
\caption{Box plots with average values (+) of traveled distances and number of transfers for trips made with and without clusters of bus stops .}
\label{fig:comparativo}
\end{figure}

The results presented in this section are based on an origin-destination (OD) survey by IPPUC in Curitiba \cite{IPPUC}. Given an OD pair, the closest bus stops from the origin and destination are identified using a search distance of 600~m. Then, a short-path algorithm computes a feasible bus trip in the network and obtains its distance traveled and the number of transfers between bus lines. When a transfer is made in a cluster, an additional walking distance computed between the bus stops is considered. Because several trips can be found for a single OD pair, Yen's algorithm \cite{Yen1971} computes K-shortest paths with $K=30$. In other words, the number of shortest paths is limited to the top 30 alternatives. The results for distance traveled and the number of transfers are shown in \textbf{Figure~\ref{fig:comparativo}}, with and without clusters. 

According to \textbf{Figure~\ref{fig:comparativo_custo}}, the average distance of 22.4~km traveled in the original network (without clusters) is longer than the average distance of 12.3~km using clusters. The opposite is observed concerning the number of transfers as shown in \textbf{Figure~\ref{fig:comparativo_custo_baldeacao}}. The average value is two transfers without clusters, while the average number of transfers is four with clusters. Therefore, the results suggest that trip distances with bus clusters decrease by almost half at the expense of twice the number of transfers on average.

\section{Conclusion}\label{sec:conclusao}

This work proposed data-driven approaches for detecting bus itineraries from GPS data 
and integrating bus transit in space and time. This spatiotemporal integration allows passengers to switch bus lines with a single fare by defining ``virtual terminals'' in specific walking distance areas where transfers can occur during a limited timeframe. 

The first algorithm for itinerary detection outcomes valid itineraries in most cases -- improving other proposals in the state-of-the-art.
The results show an increase from 68.83\% to 99.33\% in itinerary traceability gain when compared with a method that uses bus timetables. This result increases valid data in the database, preventing them from being discarded due to not being associated with any itinerary.

The second algorithm for bus stop clustering groups bus stops in walking distance areas for establishing ``virtual terminals'' where bus transfers can occur outside traditional physical terminals.
An analysis using real-world origin-destination trips in Curitiba revealed that our approach could potentially reduce travel distances significantly.
The average distance of 22.4~km traveled in the transit network without clusters is reduced to 12.3~km with clusters. However, it increases the number of transfers by two on average.


The results are limited regarding time 
estimated at bus stops because road traffic conditions should not affect them significantly when bus stops are located at short distances from each other. Another important limitation is using the correlation between bus time series to measure transfer times. A strong correlation 
means that buses are more 
likely to meet each other at bus stops of the same cluster.

Our contribution 
can enhance the efficiency of bus transit 
and even attract more people to public transport.
Several future works could be done in this direction. For instance, it may be interesting to consider arrival times for evaluating and selecting routes with better synchronization and also allowing travel time to be computed. 

\section*{Acknowledgements}
The authors thank Coordenação de Aperfeiçoamento de Pessoal de Nível Superior - Brasil (CAPES), Project Smart City Concepts in Curitiba, IPPUC, and Curitiba City Hall. JB thanks HUB de Inteligência Artificial e Arquiteturas Cognitivas - Brasil (H.IAAC), and Serviço Nacional de Aprendizagem Industrial - Brasil (SENAI).

\section*{Authors’ Contributions}
JB and AP performed the experiments. JB, TS, AM, and RL helped in the conceptualization of the study and writing of the manuscript. JB is the main contributor and writer of this manuscript. All authors read and approved the final manuscript.

\section*{Availability of data and materials}
The datasets generated and/or analyzed during the current study are available in \url{https://github.com/jcnborges/busanalysis.git}.

\bibliographystyle{apalike-sol}
\bibliography{references}  

\begin{thebibliography}{}

\bibitem[Bona {\em et~al}., 2016]{Bona2016}
Bona, A. A.~D., Fonseca, K.~V., Rosa, M.~O., Lüders, R., and Delgado, M.~R. (2016).
\newblock Analysis of public bus transportation of a {B}razilian city based on the theory of complex networks using the p-space.
\newblock {\em Mathematical Problems in Engineering}, 2016. DOI: \href{http://dx.doi.org/10.1155/2016/3898762}{10.1155/2016/3898762}.

\bibitem[Borges {\em et~al}., 2023]{courb23Julio}
Borges, J., Lüders, R., Silva, T., and Munaretto, A. (2023).
\newblock Algoritmo para detecção de itinerários do transporte público usando dados de gps dos Ônibus.
\newblock In {\em Anais do VII Workshop de Computação Urbana}, pages 1--14, Porto Alegre, RS, Brasil. SBC. DOI: \href{http://dx.doi.org/10.5753/courb.2023.739}{10.5753/courb.2023.739}.

\bibitem[Caminha {\em et~al}., 2018]{Caminha2018}
Caminha, C., Furtado, V., Pinheiro, V., and Ponte, C. (2018).
\newblock Graph mining for the detection of overcrowding and waste of resources in public transport.
\newblock {\em Journal of Internet Services and Applications}, 9:22. DOI: \href{http://dx.doi.org/10.1186/s13174-018-0094-3}{10.1186/s13174-018-0094-3}.

\bibitem[Chawuthai {\em et~al}., 2023]{Chawuthai2023}
Chawuthai, R., Sumalee, A., and Threepak, T. (2023).
\newblock {GPS} data analytics for the assessment of public city bus transportation service quality in {B}angkok.
\newblock {\em Sustainability}, 15(7). DOI: \href{http://dx.doi.org/10.3390/su15075618}{10.3390/su15075618}.

\bibitem[Curzel {\em et~al}., 2019]{Curzel2019}
Curzel, J.~L., Lüders, R., Fonseca, K.~V., and Rosa, M.~O. (2019).
\newblock Temporal performance analysis of bus transportation using link streams.
\newblock {\em Mathematical Problems in Engineering}, 2019. DOI: \href{http://dx.doi.org/10.1155/2019/6139379}{10.1155/2019/6139379}.

\bibitem[Desai {\em et~al}., 2022]{Desai2022}
Desai, S., Suthar, R., Yadav, V., Ankar, V., and Gupta, V. (2022).
\newblock Smart bus fleet management system using {IoT}.
\newblock In {\em 2022 Fourth International Conference on Emerging Research in Electronics, Computer Science and Technology (ICERECT)}, pages 01--06. DOI: \href{http://dx.doi.org/10.1109/ICERECT56837.2022.10059646}{10.1109/ICERECT56837.2022.10059646}.

\bibitem[Gallotti and Barthelemy, 2015a]{Gallotti2015b}
Gallotti, R. and Barthelemy, M. (2015a).
\newblock Anatomy and efficiency of urban multimodal mobility.
\newblock {\em Scientific Reports}, 4:6911. DOI: \href{http://dx.doi.org/10.1038/srep06911}{10.1038/srep06911}.

\bibitem[Gallotti and Barthelemy, 2015b]{Gallotti2015}
Gallotti, R. and Barthelemy, M. (2015b).
\newblock The multilayer temporal network of public transport in {G}reat {B}ritain.
\newblock {\em Scientific Data}, 2:140056. DOI: \href{http://dx.doi.org/10.1038/sdata.2014.56}{10.1038/sdata.2014.56}.

\bibitem[Gubert {\em et~al}., 2023]{gubert2023strategies}
Gubert, F.~R., Santin, P., Fonseca, M., Munaretto, A., and Silva, T.~H. (2023).
\newblock On strategies to help reduce contamination on public transit: a multilayer network approach.
\newblock {\em Applied Network Science}, 8(1):1--22. DOI: \href{http://dx.doi.org/10.1007/s41109-023-00562-7}{10.1007/s41109-023-00562-7}.

\bibitem[Hakeem {\em et~al}., 2022]{Hakeem2022}
Hakeem, M. F. M.~A., Sulaiman, N.~A., Kassim, M., and Isa, N.~M. (2022).
\newblock {IoT} bus monitoring system via mobile application.
\newblock In {\em 2022 IEEE International Conference on Automatic Control and Intelligent Systems (I2CACIS)}, pages 125--130. DOI: \href{http://dx.doi.org/10.1109/I2CACIS54679.2022.9815268}{10.1109/I2CACIS54679.2022.9815268}.

\bibitem[IPPUC, 2017]{IPPUC}
IPPUC (2017).
\newblock {Consolidação de Dados de Oferta, Demanda, Sistema Viário e Zoneamento: Relatório 5 - Pesquisa Origem-Destino Domiciliar}.
\newblock URL: \url{http://admsite2013.ippuc.org.br/arquivos/documentos/D536/D536_002_BR.pdf} (Last accessed: 2023-06-14).

\bibitem[Lawhead, 2015]{lawhead2015learning}
Lawhead, J. (2015).
\newblock {\em Learning geospatial analysis with Python}.
\newblock Packt Publishing Ltd, Birmingham, 2nd edition.

\bibitem[Li and Rong, 2022]{Li2022}
Li, T. and Rong, L. (2022).
\newblock Spatiotemporally complementary effect of high-speed rail network on robustness of aviation network.
\newblock {\em Transportation Research Part A: Policy and Practice}, 155:95--114. DOI: \href{http://dx.doi.org/https://doi.org/10.1016/j.tra.2021.10.020}{https://doi.org/10.1016/j.tra.2021.10.020}.

\bibitem[Ma {\em et~al}., 2022]{ma2022multi}
Ma, J., Chan, J., Rajasegarar, S., and Leckie, C. (2022).
\newblock Multi-attention graph neural networks for city-wide bus travel time estimation using limited data.
\newblock {\em Expert Systems with Applications}, 202:117057. DOI: \href{http://dx.doi.org/10.1016/j.eswa.2022.117057}{10.1016/j.eswa.2022.117057}.

\bibitem[Ma {\em et~al}., 2019]{ma2019bus}
Ma, J., Chan, J., Ristanoski, G., Rajasegarar, S., and Leckie, C. (2019).
\newblock Bus travel time prediction with real-time traffic information.
\newblock {\em Transportation Research Part C: Emerging Technologies}, 105:536--549. DOI: \href{http://dx.doi.org/10.1016/j.trc.2019.06.008}{10.1016/j.trc.2019.06.008}.

\bibitem[Maduako {\em et~al}., 2019]{Maduako2019c}
Maduako, I.~D., Wachowicz, M., and Hanson, T. (2019).
\newblock Transit performance assessment based on graph analytics.
\newblock {\em Transportmetrica A: Transport Science}, 15(2):1382--1401. DOI: \href{http://dx.doi.org/10.1080/23249935.2019.1596991}{10.1080/23249935.2019.1596991}.

\bibitem[Martins {\em et~al}., 2022]{Martins2022}
Martins, T., Kozievitch, N., Gadda, T., Rosa, M., and Gutierrez, M. (2022).
\newblock Map matching: Uma análise de dados streaming de trajetórias de {GPS} no transporte público.
\newblock In {\em Temas Emergentes: Cidades Inteligentes (XVIII SBSI)}, pages 294--301. SBC. DOI: \href{http://dx.doi.org/10.5753/sbsi\_estendido.2022.221647}{10.5753/sbsi\_estendido.2022.221647}.

\bibitem[Motta {\em et~al}., 2013]{Motta2013}
Motta, R.~A., Silva, P. C. M.~D., and Santos, M. P. D.~S. (2013).
\newblock Crisis of public transport by bus in developing countries: a case study from brazil.
\newblock {\em International Journal of Sustainable Development and Planning}, 8:348--361. DOI: \href{http://dx.doi.org/10.2495/SDP-V8-N3-348-361}{10.2495/SDP-V8-N3-348-361}.

\bibitem[Mulerikkal {\em et~al}., 2022]{mulerikkal2022performance}
Mulerikkal, J., Thandassery, S., Rejathalal, V., and Kunnamkody, D. M.~D. (2022).
\newblock Performance improvement for metro passenger flow forecast using spatio-temporal deep neural network.
\newblock {\em Neural Computing and Applications}, pages 1--12. DOI: \href{http://dx.doi.org/10.1007/s00521-021-06522-5}{10.1007/s00521-021-06522-5}.

\bibitem[Panigrahi, 2014]{panigrahi2014computing}
Panigrahi, N. (2014).
\newblock {\em Computing in geographic information systems}.
\newblock CRC Press, Boca Raton, Florida, 1st edition.

\bibitem[Park {\em et~al}., 2020]{Park2020}
Park, Y., Mount, J., Liu, L., Xiao, N., and Miller, H.~J. (2020).
\newblock Assessing public transit performance using real-time data: spatiotemporal patterns of bus operation delays in columbus, ohio, usa.
\newblock {\em International Journal of Geographical Information Science}, 34:367--392. DOI: \href{http://dx.doi.org/10.1080/13658816.2019.1608997}{10.1080/13658816.2019.1608997}.

\bibitem[Peixoto {\em et~al}., 2020]{Peixoto2020}
Peixoto, A., Rosa, M., Lüders, R., and Fonseca, K. (2020).
\newblock Plataforma computacional para construção de um banco de dados de grafo do sistema de transporte de {C}uritiba.
\newblock In {\em IV Workshop de Computação Urbana}, pages 125--137. SBC. DOI: \href{http://dx.doi.org/10.5753/courb.2020.12358}{10.5753/courb.2020.12358}.

\bibitem[Queiroz {\em et~al}., 2019]{Queiroz2019}
Queiroz, A.~R., Santos, V., Nascimento, D., and Pires, C.~E. (2019).
\newblock Conformity analysis of {GTFS} routes and bus trajectories.
\newblock In {\em XXXIV Simpósio Brasileiro de Banco de Dados}, pages 199--204. SBC. DOI: \href{http://dx.doi.org/10.5753/sbbd.2019.8823}{10.5753/sbbd.2019.8823}.

\bibitem[Rodrigues {\em et~al}., 2017]{Rodrigues2017}
Rodrigues, D.~O., Boukerche, A., Silva, T.~H., Loureiro, A.~A., and Villas, L.~A. (2017).
\newblock {SMAFramework: Urban Data Integration Framework for Mobility Analysis in Smart Cities}.
\newblock In {\em Proceedings of the 20th ACM International Conference on Modelling, Analysis and Simulation of Wireless and Mobile Systems}, MSWiM '17, page 227–236, New York, NY, USA. Association for Computing Machinery. DOI: \href{http://dx.doi.org/10.1145/3127540.3127569}{10.1145/3127540.3127569}.

\bibitem[Rosa {\em et~al}., 2020]{Rosa2020}
Rosa, M.~O., Fonseca, K. V.~O., Kozievitch, N.~P., De-Bona, A.~A., Curzel, J.~L., Pando, L.~U., Prestes, O.~M., and L{\"u}ders, R. (2020).
\newblock {\em Advances on Urban Mobility Using Innovative Data-Driven Models}, pages 1--38.
\newblock Springer International Publishing, Cham. DOI: \href{http://dx.doi.org/10.1007/978-3-030-15145-4\_57-1}{10.1007/978-3-030-15145-4\_57-1}.

\bibitem[Sadeghian {\em et~al}., 2021]{Sadeghian2021}
Sadeghian, P., Håkansson, J., and Zhao, X. (2021).
\newblock Review and evaluation of methods in transport mode detection based on {GPS} tracking data.
\newblock {\em Journal of Traffic and Transportation Engineering (English Edition)}, 8(4):467--482. DOI: \href{http://dx.doi.org/10.1016/j.jtte.2021.04.004}{10.1016/j.jtte.2021.04.004}.

\bibitem[Santin {\em et~al}., 2020]{Santin2020b}
Santin, P., Gubert, F.~R., Fonseca, M., Munaretto, A., and Silva, T.~H. (2020).
\newblock Characterization of public transit mobility patterns of different economic classes.
\newblock {\em Sustainability}, 12(22). DOI: \href{http://dx.doi.org/10.3390/su12229603}{10.3390/su12229603}.

\bibitem[Singla and Bhatia, 2015]{Singla2015}
Singla, L. and Bhatia, P. (2015).
\newblock {GPS} based bus tracking system.
\newblock In {\em 2015 International Conference on Computer, Communication and Control (IC4)}, pages 1--6. DOI: \href{http://dx.doi.org/10.1109/IC4.2015.7375712}{10.1109/IC4.2015.7375712}.

\bibitem[Sridevi {\em et~al}., 2017]{Sridevi2017}
Sridevi, K., Jeevitha, A., Kavitha, K., Sathya, K., and Narmadha, K. (2017).
\newblock Smart bus tracking and management system using {IoT}.
\newblock {\em Asian Journal of Applied Science and Technology (AJAST)}, 1(2).
\newblock Available at SSRN: \url{https://ssrn.com/abstract=2941150}.

\bibitem[URBS, 2022a]{URBS_RIT}
URBS (2022a).
\newblock Características da rede integrada de transporte.
\newblock URL: \url{https://www.urbs.curitiba.pr.gov.br/transporte/rede-integrada-de-transporte} (Last accessed: 2023-03-27).

\bibitem[URBS, 2022b]{PDA_URBS_API}
URBS (2022b).
\newblock Web-service: Dados públicos da rede integrada do transporte coletivo de {C}uritiba.
\newblock URL: \url{https://www.curitiba.pr.gov.br/dadosabertos/busca/?grupo=8} (Last accessed: 2023-03-27).

\bibitem[Vila {\em et~al}., 2016]{vila2016urban}
Vila, J. J.~R., Kozievitch, N.~P., Gadda, T.~M., Fonseca, K., Rosa, M.~O., Gomes-Jr, L.~C., and Akbar, M. (2016).
\newblock Urban mobility challenges--an exploratory analysis of public transportation data in {C}uritiba.
\newblock {\em Revista de Inform{\'a}tica Aplicada}, 12(1).
\newblock Available at RIA: \url{https://seer.uscs.edu.br/index.php/revista_informatica_aplicada/article/view/6905/2996}.

\bibitem[War {\em et~al}., 2022]{War2022}
War, M.~M., Rakhra, M., and Singh, D. (2022).
\newblock Review on application based bus tracking system.
\newblock In {\em 2022 5th International Conference on Contemporary Computing and Informatics (IC3I)}, pages 876--880. DOI: \href{http://dx.doi.org/10.1109/IC3I56241.2022.10072449}{10.1109/IC3I56241.2022.10072449}.

\bibitem[Wehmuth {\em et~al}., 2018]{Wehmuth2018}
Wehmuth, K., Costa, B., Bechara, J.~V., and Ziviani, A. (2018).
\newblock A multilayer and time-varying structural analysis of the {B}razilian air transportation network.
\newblock In {\em Latin America Data Science Workshop}, volume 2170 of {\em {CEUR} Workshop Proceedings}, pages 57--64.
\newblock {Available at LADaS:} \url{https://ceur-ws.org/Vol-2170/paper8.pdf}.

\bibitem[Welch and Widita, 2019]{Welch2019}
Welch, T.~F. and Widita, A. (2019).
\newblock Big data in public transportation: a review of sources and methods.
\newblock {\em Transport Reviews}, 39(6):795--818. DOI: \href{http://dx.doi.org/10.1080/01441647.2019.1616849}{10.1080/01441647.2019.1616849}.

\bibitem[Yen, 1971]{Yen1971}
Yen, J.~Y. (1971).
\newblock Finding the k shortest loopless paths in a network.
\newblock {\em Management Science}, 17:712--716. DOI: \href{http://dx.doi.org/10.1287/mnsc.17.11.712}{10.1287/mnsc.17.11.712}.

\bibitem[Yin {\em et~al}., 2014]{Yin2014}
Yin, L., Hu, J., Huang, L., Zhang, F., and Ren, P. (2014).
\newblock Detecting illegal pickups of intercity buses from their {GPS} traces.
\newblock In {\em 17th International IEEE Conference on Intelligent Transportation Systems (ITSC)}, pages 2162--2167. DOI: \href{http://dx.doi.org/10.1109/ITSC.2014.6958023}{10.1109/ITSC.2014.6958023}.

\bibitem[Yu {\em et~al}., 2020]{yu2020policy}
Yu, Q., Li, W., Yang, D., and Xie, Y. (2020).
\newblock Policy zoning for efficient land utilization based on spatio-temporal integration between the bicycle-sharing service and the metro transit.
\newblock {\em Sustainability}, 13(1):141. DOI: \href{http://dx.doi.org/10.3390/su13010141}{10.3390/su13010141}.

\bibitem[Zhang {\em et~al}., 2021]{Zhang2021}
Zhang, H., Liu, Y., Shi, B., Jia, J., Wang, W., and Zhao, X. (2021).
\newblock Analysis of spatial-temporal characteristics of operations in public transport networks based on multisource data.
\newblock {\em Journal of Advanced Transportation}, 2021:1--15. DOI: \href{http://dx.doi.org/10.1155/2021/6937228}{10.1155/2021/6937228}.

\bibitem[Zhao {\em et~al}., 2023]{zhao2023developing}
Zhao, T., Huang, Z., Tu, W., Biljecki, F., and Chen, L. (2023).
\newblock Developing a multiview spatiotemporal model based on deep graph neural networks to predict the travel demand by bus.
\newblock {\em International Journal of Geographical Information Science}, pages 1--27. DOI: \href{http://dx.doi.org/10.1080/13658816.2023.2203218}{10.1080/13658816.2023.2203218}.

\end{thebibliography}


\end{document}